\documentclass[11pt]{amsart}

\usepackage{amssymb}
\usepackage{epsfig}
\usepackage{amsmath}
\usepackage{subcaption}
\usepackage[section]{placeins}
\usepackage{mathrsfs}
\usepackage{graphicx}
\usepackage{ifthen}
\newcommand{\includegraphicssafe}[2][]{\IfFileExists{#2}{\includegraphicssafe[#1]{#2}}{\fbox{Missing figure: #2}}}

\graphicspath{{./}{Figures/}}
\usepackage[notrig]{physics}
\usepackage{units}
\usepackage{pgffor}
\usepackage{changes}
\definechangesauthor[name=Ignacio,color=teal]{IR}
\definechangesauthor[name=Michael,color=orange]{MO}

\usepackage{color}
\usepackage{pgffor}

\newtheorem{thm}{Theorem}[section]

\theoremstyle{definition}
\newtheorem{definition}[thm]{Definition}
\newtheorem{example}[thm]{Example}

\theoremstyle{remark}
\newtheorem{remark}[thm]{Remark}

\begin{document}

\title[Optimal history variables]{Identification of optimal history variables and corresponding hereditary laws in linear viscoelasticity}

\author{I.~Romero$^{1,2}$ and M.~Ortiz$^{3,4}$}

\address
{
${}^1$Universidad Polit\'ecnica de Madrid,
Mechanical Engineering Department, 28006 Madrid, Spain
\\
${}^2$IMDEA Materials Institute, 
28906 Madrid, Spain
\\
${}^3$California Institute of Technology,
Engineering and Applied Science Division, Pasadena CA, 91125, USA
\\
${}^{4}$Centre Internacional de Métodes Numerics en Enginyeria (CIMNE), Universitat Politècnica de Catalunya, Jordi Girona 1, 08034 Barcelona, Spain. 
}

\email{\texttt{ignacio.romero@upm.es}, \texttt{ortiz@caltech.edu}}

%

\begin{abstract}
We develop an operator-theoretic formulation of hereditary constitutive models and characterize optimal finite-rank internal-variable approximations in the sense of Kolmogorov $N$-widths. The history operator is shown to be compact under natural assumptions on the relaxation kernel, thereby admitting optimal low-rank approximations. The resulting reduced models inherit thermodynamic consistency, stability, and provable approximation bounds. An analysis clarifies the structural relation between hereditary representations and internal-variable theories and provides a rigorous basis for reduced-order modeling in computational mechanics. Selected numerical examples showcase optimal convergence of approximations with respect to rank and sampling. 
\end{abstract}

\maketitle


\section{Introduction}

The availability of big material data sets, made possible by advances in experimental and computational science (see, e.~g., \cite{Sutton:2009, Buljac:2018, Schleder:2019, Bernier:2020, Wang:2024, Li:2019, Jin:2022}), has given rise to a desire to forge a closer nexus between material data and the predictions they enable. Remarkable experimental advances, including Dynamic Mechanical Analysis (DMA) \cite{Menard:2002}, nanoindentation \cite{Herbert:2008, Herbert:2009}, Dynamic Shear Testing (DST) \cite{Arbogast:1998, Bayly:2008}, Fourier Transform Rheology \cite{Wilhelm:1998}, high-throughput experiments \cite{Breedveld2003MicrorheologyHTS, Schultz2011HTRMicrofluidic, Zhang2023HTCHydrogels}, and others, have enabled researchers to characterize viscoelastic properties with high precision. In addition, the ability to interrogate detailed representative volume elements (RVEs) of microstructured and architectured materials along arbitrary strain paths computationally, with well-characterized materials at the microscale and with high numerical fidelity, also has demonstrated a potential for generating large material data sets \cite{Reppel2014PolyureaViscoelastic, Krushynska2016JMPS, Buchen2021EPDMViscoelastic, Fischbach2023AgingCreepPLA, Abuzayed2024TimeDomainHomog, Liupekevicius2025CMA}. 

Two main paradigms have emerged in response to the abundance of material data, loosely corresponding to supervised and unsupervised methods in machine learning: {\sl Model-free} approaches, in which material set data is combined directly with field equations to effect predictions of quantities of interest \cite{Kirchdoerfer:2016, Kirchdoerfer:2017, Conti:2018, Eggersmann:2019, Salahshoor:2023}; and {\sl model-based} approaches, in which the connection between material data and predictions is effected through the intermediate step of identifying a material law from the data \cite{Liu:2022, Bhattacharya:2023, Liu:2023, Weinberg:2023, Asad:2023, Marino:2023, Ghane:2024, Akerson:2024}. 

In the context of the second paradigm, it has been long recognized that material identification from empirical data may be regarded as an inverse problem (see, e.~g., the pioneering work of Bui \cite{Bui:1993}). For the most part, the classical work is concerned mainly with the identification of parameters in a given class of models, e.~g., polynomial expansions \cite{Pipkin:1968, pipkin1961a, pipkin1964a, rivlin1965a} or Prony series \cite{Prony:1795, Qvale:2004, Knauss:2007, Zhao:2007}, in contrast to the more challenging problem of identifying the functional form of the hereditary law itself (see, e.~g., \cite{Martins:2018} and references therein). Neural networks and machine learning have supplied a new and efficient means of representing material laws and fitting them by regression to big data sets, causing an extensive reevaluation of the field \cite{Liu:2022, Bhattacharya:2023, Liu:2023, Asad:2023, Marino:2023, Ghane:2024, Akerson:2024}.

Whereas these representations are convenient and efficient in practice, they are based on an {\sl a priori} assumption of a particular parameterized form of the hereditary law, which begs the question of what is the {\sl best}, or {\sl optimal}, representation of a given viscoelastic material, or a class of viscoelastic materials, by finite-rank hereditary operators. This problem falls squarely within the theory of {\sl $N$-widths} \cite{Pinkus:1985}, and was solved by Schmidt as early as 1907 \cite{Schmidt:1907}, with further seminal contributions by such giants as A.~Kolmogoroff \cite{Kolmogoroff:1936}, I.~M.~Gel'fand \cite{Gelfand:1959}, V.~M.~Tikhomirov \cite{Tikhomirov:1960}, and others. The theory extends to the case in which the hereditary law is not known exactly but it is only known to belong to a certain class of hereditary laws, e.~g., as defined by an experimental data set. 

The appeal of the theory of $N$-widths is that it supplies subspaces of histories of given dimension resulting in the best possible approximation of a class of hereditary laws. We note that the approximation of hereditary laws by finite-rank operators is in fact equivalent to the formulation of viscoelastic models in terms of a finite number of {\sl history} or {\sl internal variables}. The theory of $N$-widths thus also answers the question of what is the best choice of history or internal variables for purposes of representing a given class of linear viscoelastic materials, a problem lucidly formulated in \cite{Liu:2023, Bhattacharya:2023}. 

In this paper, we present an efficient numerical implementation based on the theory of $N$-widths
that supplies optimal low-rank hereditary representations of viscoelastic behavior of materials. The
assumption is that the material behavior can be interrogated exactly, or with controlled error,
along arbitrary strain paths, either experimentally or computationally, but such interrogation is
expensive and cannot be performed on-the-fly as part of large-scale simulations. The
characterization of the material must therefore be performed offline and prior to the simulations.
However, this situation raises a number of theoretical and practical questions, to wit: Which is the
optimal collection of strain paths for characterizing a particular material? What are optimal
choices of history variables and corresponding low-rank hereditary representations of the
viscoelastic behavior of the material? What are efficient implementations of the resulting
identification scheme and resulting time-stepping calculations? We address these questions using
basis representations on the natural Hilbert-space structure of spaces of histories. We demonstrate the efficiency and optimality of the scheme, relative to {\sl ad hoc} representations, and the practicality of the resulting viscoelastic material laws with the aid of selected numerical tests. 

\section{Linear viscoelasticity}

The axiomatic and empirical basis of linear viscoelasticity is well established and the subject of an extensive literature (e.~g., \cite{coleman1961b, Gurtin:1962, fisher1973a}). We consider isothermal processes throughout and omit any and all dependences on temperature for simplicity of notation. 

\subsection{The hereditary law}

By {\sl local strain and stress evolutions} we understand functions, denoted $\epsilon : \mathbb{R} \to E$ and $\sigma : \mathbb{R} \to F$, defined over the real time line $\mathbb{R}$ with values in a finite-dimensional linear space $E$ and its dual $F = E^*$, respectively. For $t \in \mathbb{R}$, $\epsilon(t) \in E$ and $\sigma(t) \in F$ then denote the strain and stress at time $t$ and $\sigma(t) \cdot \epsilon(t) \in \mathbb{R}$ denotes their duality pairing. 

We work throughout within the linearized kinematics framework. In this setting, the terms `strain' and `stress' refer generally to work-conjugate variables representing the local state of deformation and the local state of internal force, respectively. However, for definiteness we confine attention to three-dimensional linear viscoelasticity and identify $E = \mathbb{R}^{3\times 3}_{\rm sym}$, the linear space of $3\times 3$ symmetric matrices. 

An axiomatic foundation for linear viscoelasticity can be built on the Boltzmann {\sl superposition principle} \cite{Boltzmann:1874}. Appealing, in addition, to fundamental principles such as causality, the dissipation inequality and reciprocity, as well as additional assumptions such as time-shift invariance, integrability of the kernel and no instantaneous viscosity, it follows \cite{coleman1961b,Gurtin:1962} that the most general relation between strain and stress evolutions is the hereditary law
\begin{equation} \label{29a8Zn}
  \sigma(t)
  =
  \mathbb{C} \, \epsilon(t)
  -
  \int_{-\infty}^t
    \mathbb{K}(t-s)
     \, \epsilon(s)
  \, ds 
  :=
  \mathbb{C} \, \epsilon(t)
  -
  (\mathbb{K} * \epsilon)(t) ,
\end{equation}
where $\mathbb{C} \in L(E,F)$ is the elasticity tensor, $\mathbb{K} : \mathbb{R} \to L(E,F)$ is the hereditary kernel and $(*)$ is the convolution operator. In addition, causality requires the hereditary kernel to be one-sided, i.~e.,
\begin{equation} \label{l3S3XT}
  \mathbb{K}(t) = 0 ,
  \quad
  t < 0 ,
\end{equation}
whereas reciprocity and the dissipation inequality in turn require
\begin{subequations} \label{R4H4Kc} 
\begin{align}
  &
  \mathbb{C}^T = \mathbb{C} ,
  \quad
  \mathbb{C} \geq 0 ,
  \\ &
  \mathbb{K}^T(\tau) = \mathbb{K}(\tau) ,
  \quad
  \mathbb{K}(\tau) \geq 0 ,
  \quad
  \tau \geq 0 .
\end{align}
\end{subequations}
We take the hereditary law (\ref{29a8Zn}), and the physical requirements (\ref{l3S3XT}) and (\ref{R4H4Kc}), which are assumed to be in force throughout, as point of departure for all further developments.

\subsection{Relaxation spectrum} \label{ZRp3LD}

A broad class of hereditary kernels can be formulated using spectral representations \cite[\S4]{tschoegl1989}, of which Prony series \cite{Prony:1795} are a special case. In the numerical examples that follow, we shall use such representations to characterize viscoelastic behavior at the microscale, the macroscopic behavior then resulting from a representative volume element (RVE) calculation \cite{Abuzayed2024TimeDomainHomog, Liupekevicius2025CMA}. 

The notion that the rheology of materials arises from the superposition of internal mechanisms, each characterized by a {\sl relaxation time}, was introduced by Wiechert \cite{Wiechert:1893}, and pervades much of the theory and praxis of linear viscoelasticity. The collection of relaxation times, or {\sl relaxation spectrum}, can be finite, countable or continuous (see, \cite[\S4]{tschoegl1989}; also Kestin and Rice \cite{Kestin:1970} for a critical review). 

This connection can be rendered explicit by writing a general hereditary kernel in the form
\begin{equation} \label{eqZ2Sh}
  \mathbb{K}(\tau)
  =
  \int_{\lambda_0}^{+\infty}
    (\lambda-\lambda_0) \, 
    {\rm e}^{-\lambda \tau}
  \, d\nu(\lambda) ,
  \quad
  \tau \geq 0 ,
\end{equation}
where $1/\lambda$ is a generic relaxation time and $\nu$ is an $L(E,F)$-valued measure with support in $[\lambda_0,+\infty)$, or {\sl relaxation measure}, with $1/\lambda_0$ a cutoff relaxation time, possibly infinite. The support of $\nu$ is the {\sl relaxation spectrum}. Alternative spectral representations can be based on the Laplace transform~\cite[\S4]{tschoegl1989}. 

Conditions on the relaxation measure resulting in well-behaved hereditary kernels are presented in \cite{Gurtin:1962,Ortiz:2025}. For instance, if the hereditary kernel is bounded, then it suffices for the relaxation measure to have bounded total mass, i.~e.,
\begin{equation} 
  | \nu |
  =
  \int_{\lambda_0}^{+\infty}
  d\nu(\lambda) 
  \leq
  C < +\infty, 
\end{equation}
for some positive constant $C > 0$. The use of relaxation measures extends the classical treatment based on continuous densities \cite[\S4]{tschoegl1989} and unifies the treatment of discrete and continuous spectra. 

\begin{example}[Maxwell-Wiechert model] \label{t3QNXc} {\rm A prominent example of a spectral representation is furnished by the Maxwell-Wiechert model \cite{Wiechert:1893, Wiechert:1899} and attendant Prony series \cite{Prony:1795}. The Maxwell-Wiechert model can be recast in the form (\ref{eqZ2Sh}) by choosing a relaxation measure 
\begin{equation}
  \nu_{ijkl}
  =
  \Big( \nu_L - \frac{2}{3} \nu_M \Big)\delta_{ij}\delta_{kl}
  +
  \nu_M (\delta_{ik}\delta_{jl} + \delta_{il}\delta_{jk}) ,
\end{equation}
with
\begin{equation} \label{yo7cHo}
  \nu_L({\lambda})
  =
  \sum_{i=1}^l
  \frac{L_i}{\lambda-\lambda_0}
  \delta_{\lambda_i} ,
  \quad
  \nu_M({\lambda})
  =
  \sum_{i=1}^m
  \frac{M_i}{\lambda-\lambda_0}
  \delta_{\mu_i} ,
\end{equation}
$0< \lambda_0 < \lambda_i$, $i=1,\dots,l$, $0 < \lambda_0 < \mu_i$, $i=1,\dots,m$, and $\delta_{\lambda_i}$ and $\delta_{\mu_i}$ Dirac measures centered at $\lambda_i$ and $\mu_i$, respectively. The universal approximation property of Prony series in the class of spectral models is discussed in \cite{Ortiz:2025,tschoegl1989}.
} \hfill$\square$
\end{example}

\subsection{Representative volume element representations} \label{NOtUv7}

In multiscale representations of material behavior, the macroscopic material law represents the effective behavior of a representative volume element (RVE). If the material behavior is viscoelastic at the microscale, then it is readily shown that the material behavior at the macroscale is also viscoelastic and characterized by an effective hereditary law. For simplicity, we assume that the RVE is discrete, e.~g., a viscoelastic metamaterial \cite{Krushynska2016JMPS} or a finite-element discretization of a viscoelastic solid \cite{Abuzayed2024TimeDomainHomog}. We label by $e = 1,\dots, m$ the material points in the RVE. The governing equations are then
\begin{subequations}
\begin{align}
  &  \label{eKISpO}
  \sum_{e=1}^m w_e B^{eT} (\sigma_e(t) - \bar{\sigma}(t))= 0 ,
  \\ &  \label{yb774v}
  \epsilon_e(t) = B_e u(t) ,
  \\ &  \label{NIqzBv}
  \sigma_e(t) 
  =
  \mathbb{C}_e \, \epsilon_e(t)
  -
  \int_{-\infty}^t
    \mathbb{K}_e(t-s)
     \, \epsilon_e(s)
  \, ds ,  
\end{align}
\end{subequations} 
where (\ref{eKISpO}) are the equations of equilibrium, (\ref{yb774v}) the compatibility equations, (\ref{NIqzBv}) the local hereditary laws, assumed known, $u(t) \in \mathbb{R}^n$ is a displacement array, possibly periodic, $(w_e)_{e=1}^m$ are local weights, $(\epsilon_e(t))_{e=1}^m$ are the local strains, $(\sigma_e(t))_{e=1}^m$ are the local stresses, $(\mathbb{C}_e)_{e=1}^m$ are the local elastic moduli and $(\mathbb{K}_e(\tau))_{e=1}^m$ are the local hereditary kernels. In addition, $\bar{\sigma}(t) \in F$ is a macroscopic stress determined by a macroscopic strain constraint
\begin{equation}
  \sum_{e=1}^m w_e \epsilon_e(t) 
  = 
  \Big( \sum_{e=1}^m w_e \Big)
  \bar{\epsilon}(t) ,
\end{equation}
with $\bar{\epsilon}(t) \in E$ given. A general Laplace transform of the RVE equations using the identity
\begin{equation}
  \mathcal{L} 
  \left\{
  \mathbb{K} * \epsilon
  \right\}({s})
  = 
  \tilde{\mathbb{K}}({s})\, \tilde{\epsilon}({s}),
\end{equation}
under the assumptions that $\mathbb{K}(t)=0$ and $\epsilon(t)=0$ for $t<0$, yields
\begin{subequations}
\begin{align}
  &  
  \sum_{e=1}^m w_e B_e^T 
    (\tilde{\sigma}_e({s}) - \tilde{\bar{\sigma}}({s})) = 0 ,
  \\ &  
  \tilde{\epsilon}_e({s}) = B_e \tilde{u}({s}) ,
  \\ &  
  \tilde{\sigma}_e({s}) 
  =
  (\mathbb{C}_e - \tilde{\mathbb{K}}_e({s}) ) 
  \tilde{\epsilon}_e({s}) ,
  \\ & 
  \sum_{e=1}^m w_e \tilde{\epsilon}_e({s}) 
  = 
  \Big( \sum_{e=1}^m w_e \Big)
  \tilde{\bar{\epsilon}}({s}) ,
\end{align}
\end{subequations} 
where a superposed $( \, \tilde{} \, )$ denotes Laplace transform and ${s}$ is the Laplace variable. Eliminating $(\epsilon_e(t))_{e=1}^m$ and $(\sigma_e(t))_{e=1}^m$, we obtain
\begin{subequations}
\begin{align}
  &
  \sum_{e=1}^m w_e B^{eT} 
  (\mathbb{C}_e - \tilde{\mathbb{K}}_e({s}) ) 
  B_e \tilde{u}({s}) 
  = 
  \Big( \sum_{e=1}^m w_e B_e \Big)^T \tilde{\bar{\sigma}}({s}) ,
  \\ &
  \Big( \sum_{e=1}^m w_e B_e \Big) \tilde{u}({s}) 
  = 
  \Big( \sum_{e=1}^m w_e \Big)
  \tilde{\bar{\epsilon}}({s})
\end{align}
\end{subequations}
or, in matrix form,
\begin{subequations}
\begin{align}
  &  \label{GTyHDO}
  W B^T (\mathbb{C} - \tilde{\mathbb{K}}({s}) ) B \tilde{u}({s}) ,
  = 
  C^T \tilde{\bar{\sigma}}({s}) ,
  \\ &  \label{qxB5iT}
  C \tilde{u}({s}) = \tilde{\bar{\epsilon}}({s}) .
\end{align}
\end{subequations}
where we write $W = {\rm diag}(w_1,\dots,w_m)/\Big( \sum_{e=1}^m w_e \Big)$, $B^T = (B_1^T | \dots |B_m^T)$, $\mathbb{C} = ( \mathbb{C}_1,\dots,\mathbb{C}_m)$, $\mathbb{K} = ( \mathbb{K}_1,\dots,\mathbb{K}_m)$ and
\begin{equation}
  C = \Big(\sum_{e=1}^m w_e B_e\Big)/\Big( \sum_{e=1}^m w_e \Big) .
\end{equation}
Solving (\ref{GTyHDO}) for the displacements yields
\begin{equation}
  \tilde{u}({s})
  =
  \Big(
    W B^T (\mathbb{C} - \tilde{\mathbb{K}}({s}) ) B
  \Big)^{-1} C^T \tilde{\bar{\sigma}}({s}) ,
\end{equation}
and inserting the result into (\ref{qxB5iT}) gives the relation
\begin{equation}
  C 
  \Big(
    W B^T (\mathbb{C} - \tilde{\mathbb{K}}({s}) ) B
  \Big)^{-1} C^T \tilde{\bar{\sigma}}({s})
  = 
  \tilde{\bar{\epsilon}}({s}) ,
\end{equation}
Solving for the average stresses, finally gives
\begin{equation} \label{A8bjcL}
  \tilde{\bar{\sigma}}({s}) 
  = 
  (\bar{\mathbb{C}} - \tilde{\bar{\mathbb{K}}}({s}))
  \tilde{\bar{\epsilon}}({s}) ,
\end{equation}
where 
\begin{equation}
  \bar{\mathbb{C}}
  :=
  \Big(
    C 
    \Big(
      W B^T \mathbb{C} B
    \Big)^{-1} C^T 
  \Big)^{-1}  
\end{equation}
are effective elastic moduli and 
\begin{equation} \label{yH6SCC}
  \tilde{\bar{\mathbb{K}}}({s})
  :=
  \bar{\mathbb{C}}
  -
  \Big(
    C 
    \Big(
      W B^T (\mathbb{C} - \tilde{\mathbb{K}}({s}) ) B
    \Big)^{-1} C^T 
  \Big)^{-1}  
\end{equation}
is the effective hereditary kernel in Laplace representation. The corresponding real time representation $\bar{\mathbb{K}}(\tau)$ then follows simply by an application of the inverse Laplace transform to $\tilde{\bar{\mathbb{K}}}({s})$, whereupon (\ref{A8bjcL}) becomes
\begin{equation} 
  \bar{\sigma}(t)
  =
  \bar{\mathbb{C}} \, \bar{\epsilon}(t)
  -
  \int_{-\infty}^t
    \bar{\mathbb{K}}(t-s)
     \, \bar{\epsilon}(s)
  \, ds 
  =
  \bar{\mathbb{C}}\, \bar{\epsilon}(t)-
  (\bar{\mathbb{K}} * \bar{\epsilon})(t) .
\end{equation}
Evidently, this hereditary law is of the form (\ref{29a8Zn}), albeit expressed in terms of effective and macroscopic quantities. 

In practice, the evaluation of the effective properties, which often entails a large scale RVE calculation and Laplace transforms thereof, may be exceedingly costly. In addition, an application of Cramer's rule to (\ref{yH6SCC}) reveals that homogenization greatly increases the complexity of the hereditary kernel in general, which begs the question of efficient approximation of the macroscopic hereditary law. 

\section{History representation and approximation}

For given local stress evolution $\sigma : \mathbb{R} \to F$, the hereditary law (\ref{29a8Zn}) defines a {\sl convolution Volterra equation of the second kind} in the local strain evolution $\epsilon : \mathbb{R} \to E$ \cite{gripenberg:1990}. Necessary and sufficient stability conditions on the kernel for the local problem, i.~e., the problem of determining the strain evolution corresponding to a given stress evolution, are summarized in \cite{gripenberg:1990,Gurtin:1962}. Such stability conditions are found to be satisfied, for instance, by the Maxwell-Wiechert model of Example~\ref{t3QNXc}.

A functional framework suitable for analysis and approximation may be set forth as follows. We begin by noting that the elasticity tensor $\mathbb{C}$ can conveniently be enlisted to metrize strains and stresses, leading to the following definitions.

\begin{definition}
[Local stress and strain spaces] We define the space $M$ of local strains as the linear space $E$ metrized by $\mathbb{C}$. We define the space $N$ of local stresses as the linear space $F$ metrized by $\mathbb{C}^{-1}$. As Euclidean spaces, $N = M^*$ and $M = N^*$ and the Riesz mapping is given by Hooke's law $\sigma(t) = \mathbb{C} \epsilon(t)$. 
\end{definition}

We adopt throughout a {\sl history representation} \cite{coleman1961b, Gurtin:1962}. For a fixed material point, the {\sl past local histories} of strain and stress up to time $t$ are the functions 
\begin{equation} \label{ZrANem}
  \epsilon_t(\tau) = \epsilon(t-\tau) ,
  \quad
  \sigma_t(\tau) = \sigma(t-\tau) , 
  \quad
  \tau \geq 0 . 
\end{equation}
In terms of histories, for fixed $t$ the hereditary law (\ref{29a8Zn}) becomes
\begin{equation} \label{bgmUJR}
  \sigma_t(\tau)
  =
  \mathbb{C} \, ( I - S ) \, \epsilon_t(\tau) ,
\end{equation}
where the history operator 
\begin{equation} \label{yCn5JN}
  S \, \epsilon_t(\tau)
  =
  \int_\tau^{t} 
    \mathbb{C}^{-1} \mathbb{K}(\rho-\tau) \, 
    \epsilon_t(\rho) 
  \, d\rho ,
\end{equation}
maps local histories of strain $\epsilon_t$ to local strain histories of inelastic strain $S\epsilon_t$.

\subsection{History representation}

A suitable functional framework in which to place the operator (\ref{yCn5JN}) is the following \cite{Ortiz:2025}. To allow for---and characterize---fading memory properties, we shall measure time according to a positive, continuous, non-increasing, integrable {\sl weighting function} $w : [0,T] \to \mathbb{R}$, normalized to $w(0) = 1$, and denote by 
\begin{equation}
  d\mu(t) = w(T-t) \, dt ,
\end{equation}
the corresponding time measure. Weights, or {\sl influence functions}, were introduced by Mizel and Wang \cite{Mizel:1966} as a means of characterizing materials with fading memory. 

\begin{definition}[Spaces of local stress and strain histories] \label{D3jOho
} The space ${H}$ of local strain histories is the weighted time-dependent Lebesgue space $L^2((0,T), M, \mu)$, and the space $H^*$ of local stress evolutions is the weighted time-dependent Lebesgue space $L^2((0,T), N, \mu)$, both with the usual metrization (see, e.~g., \cite[\S5.9.2]{Evans:1998})
\begin{equation} \label{KJMHD7}
  ( \xi, \eta )
  := 
  \int_{0}^{T} 
    ( \xi(\tau), \eta(\tau) )
  \, d\mu(\tau).
\end{equation}
The Riesz mapping given by the timewise application of Hooke's law $\sigma_t(\tau) = \mathbb{C} \epsilon_t(\tau)$. 
\end{definition}

\noindent
This choice of functional framework is natural in the sense that, under physically reasonable assumptions on $\mathbb{C}$, $\mathbb{K}(\tau)$, $d\mu(\tau)$ and $\sigma(\tau)$, the Volterra equation set forth by the hereditary law (\ref{29a8Zn}) has a unique solution that depends continuously on the data \cite{gripenberg:1990, Ortiz:2025}.

In view of the {\sl Hilbert-space} structure of the spaces of histories, it is natural to resort to basis representations thereof. Let $(\varphi_k)_{k=1}^\infty$ be an orthonormal basis of ${H}$. Then, the strain histories admit the representation
\begin{equation} \label{A6y3Fr}
  \epsilon_t(\tau)
  =
  \sum_{k=1}^\infty 
  q_k(t)
  \, \varphi_k(\tau) ,
\end{equation}
where 
\begin{equation}
  \quad
  q_k(t) 
  := 
  ( \epsilon_t, \varphi_k ) 
  = 
  \int_0^{T} 
    ( \epsilon_t(\rho), \varphi_k(\rho) )
  \, d\mu(\rho),
\end{equation}
are coordinates of $\epsilon_t(\tau)$ in the basis $(\varphi_k)_{k=1}^\infty$. Likewise, introduce the representation 
\begin{equation} \label{AJ3FYj} 
  S \epsilon_t(\tau) 
  = 
  \sum_{k=1}^\infty p_k(t) 
  \, \varphi_k(\tau) ,
\end{equation}
where
\begin{equation}
  p_k(t) 
  := 
  ( S \epsilon_t, \varphi_k ) 
  = 
  \int_0^{T} 
    ( S \epsilon_t(\rho), \varphi_k(\rho) ) 
  \, d\mu(\rho),
\end{equation}
are coordinates of $S\epsilon_t(\tau)$ in the basis $(\varphi_k)_{k=1}^\infty$. Combining (\ref{A6y3Fr}) and (\ref{AJ3FYj}), we find the relation
\begin{equation} \label{XVcmZ9}
  p_k(t)
  =
  \sum_{l=1}^\infty S_{kl} \, q_l(t) ,
  \quad
  S_{kl}
  =
  ( \varphi_k , S \varphi_l ) ,
\end{equation}
provided that the series converges, which supplies a coordinate representation of the operator $S$. 

We observe from representation (\ref{A6y3Fr}) that the variables $q_k(t)$ record sufficient information to reconstruct the entire history of strain, and can therefore be regarded as {\sl history variables}. In addition, we see from (\ref{bgmUJR}) and (\ref{AJ3FYj}) that the variables $q_k(t)$, together with $\epsilon(t)$, fully characterize the instantaneous state of the material at time $t$ and, therefore, can also be interpreted as {\sl internal variables}. Internal variable representations of materials with memory date back to the work of C.~Eckart \cite{Eckart:1940, Eckart:1948}, Meixner \cite{Meixner:1953}, Biot \cite{Biot:1954} and Ziegler \cite{Ziegler:1958} and were formalized further by Coleman and Gurtin \cite{Coleman:1967} and others \cite{Rice:1971, lubliner1973a, Rice:1975} (see \cite{Horstemeyer:2010} for a historical overview). 

\begin{example}[Trigonometric-exponential basis] \label{Sg02Wo}
We wish to identify an orthonormal basis in ${H} = L^2((0,T), e^{-\lambda_0 \tau} d\tau)$, $\lambda_0 > 0$. Define the unitary map
\begin{equation} \label{U9I6tR}
  U : H \to L^2(0,T), \quad (Uf)(\tau) = e^{-\frac{\lambda_0}{2}\tau} f(\tau),
\end{equation}
with inverse
\begin{equation}
  (U^{-1}g)(\tau) = e^{\frac{\lambda_0}{2}\tau} g(\tau).
\end{equation}
Then
\begin{equation}
  ( f,g)_H
  = 
  ( Uf, Ug)_{L^2(0,T)} ,
\end{equation}
and $U$ defines an isometric isomorphism between $H$ and $L^2(0,T)$. Let $\{{e}_n\}_{n\ge1}$ be any orthonormal basis of $L^2(0,T)$. Then,
\begin{equation}
  \varphi_n(\tau) 
  := 
  U^{-1}{e}_n(\tau)
  = 
  e^{\frac{\lambda_0}{2}\tau} {e}_n(\tau),
  \quad n=1,2,\ldots
\end{equation}
forms an orthonormal basis of $H$, since for $m,n$,
\begin{equation}
  ( \varphi_m,\varphi_n)_H
  = 
  ( {e}_m,{e}_n)_{L^2(0,T)}
  = 
  \delta_{mn} ,
\end{equation}
as required. For instance, in the unweighted case the standard Fourier-type basis is
\begin{equation}
  {e}_n(\tau)
  = 
  \sqrt{\frac{2}{T}} \sin\left(\frac{n\pi\tau}{T}\right),
  \quad n=1,2,\ldots .
\end{equation}
The corresponding orthonormal basis of $H$ is
\begin{equation}
  \varphi_n(\tau)
  = 
  e^{\frac{\lambda_0}{2}\tau} \sqrt{\frac{2}{T}}
  \sin\!\left(\frac{n\pi\tau}{T}\right),
  \quad n=1,2,\ldots
\end{equation}
and similarly with cosines.
\end{example}

\subsection{The approximation property}

Operators of the form
\begin{equation} \label{zJ8aMM}
  S_N \epsilon_t(\tau) 
  = 
  \sum_{k=1}^N ( \epsilon_t, \phi_k ) \, \psi_k(\tau) ,
\end{equation} 
where $(\phi_k)_{k=1}^N$ and $(\psi_k)_{k=1}^N$ are functions in ${H}$, are said to be of {\sl finite-rank}. We wish to ascertain under what conditions hereditary operators $S$ of the form (\ref{yCn5JN}) can be approximated, in the sense of the operator norm, by sequences $(S_N)$ of finite rank operators, i.~e., 
\begin{equation}
  \lim_{N\to\infty} \| S - S_N \| = 0 . 
\end{equation}
It is well-known \cite[Cor.~6.2.]{Brezis:2010} that, in Hilbert spaces, operators have this {\sl approximation property} if and only if they are {\sl compact}, hence {\sl bounded} \cite[\S4.16]{Rudin:1991}. 

An important class of compact operators is the class of {\sl Hilbert-Schmidt operators}. We recall that $S$ is a Hilbert-Schmidt operator over a Hilbert space ${H}$ if \cite[Ex.~IX.2.19]{Conway:1990}
\begin{equation} \label{3xGVzH}
  \| S \|_{\rm HS} 
  := 
  \Big( \sum_{k=1}^\infty \| S \varphi_k \|^2 \Big)^{1/2}
  < +\infty ,
\end{equation}
where $\| S \|_{\rm HS} \geq \| S \|$ is the {\sl Hilbert-Schmidt norm} of $S$ and $(\varphi_k)_{k=1}^\infty$ is an orthonormal basis of ${H}$. It is readily verified that the definition (\ref{3xGVzH}) is independent of the choice of basis. A classical result from analysis is that Hilbert-Schmidt operators are indeed compact \cite[Ex.~4.15]{Rudin:1991}.

Compactness of Volterra operators requires boundedness of the time domain. Therefore, henceforth we restrict attention throughout to histories of finite duration $T>0$, and redefine the space of histories accordingly as ${H}$ $:=$ $L^2((0,T), M, \mu)$. Then, we have the following result \cite{Ortiz:2025}.

\begin{thm}[Hilbert-Schmidt property] \label{mCpR5D}
Assume:
\begin{itemize}
\item[i)] (Elastic stability). $\mathbb{C} \in L(\mathbb{R}^{{n}\times {n}}_{\rm sym})$, $\mathbb{C}^T = \mathbb{C}$, $\mathbb{C} > 0$.
\item[ii)] (Hilbert-Schmidt). There is a positive, continuous, non-increasing, square-integrable weighting function $w(\tau) : [0,T] \to \mathbb{R}$, normalized to $w(0) = 1$, satisfying the semigroup condition
\begin{equation}
    w(s) \ge w(s-t) w(t)\ ,
    \quad
    0\le t\le s\le T,
\end{equation}
and $\gamma > 0$ such that
\begin{equation} \label{DcjG6H}
  \int_0^{T}
    \| \mathbb{K}(\tau) \|^2
  \, w^{-1}(\tau) \, d\tau
  \leq
  \gamma^2 ,
\end{equation}
where $\| \cdot \|$ denotes the operator norm.
\end{itemize}
Then, the history operator $S$, eq.~(\ref{yCn5JN}), is Hilbert-Schmidt, hence compact, in ${H}$ and
\begin{equation} \label{4tgDYk}
  \| S \|_{\rm HS} \leq \gamma {\sqrt{T}} 
\end{equation}
\end{thm}

The proof may be found in \cite{Ortiz:2025}. We note that, indeed, the bound (\ref{4tgDYk}) degenerates if $T=+\infty$.

\begin{remark}[Encoder/decoder representation]
Finite-rank approximations such as (\ref{zJ8aMM}) may be regarded as \emph{encoder/decoder} approximations \cite{NovakWozniakowski2008, NovakWozniakowski2010, NovakWozniakowski2012}. Thus, writing
\begin{equation}
    f_N(\epsilon_t)
    =
    \{
        ( \epsilon_t, \phi_1 ),
        \dots
        ( \epsilon_t, \phi_N )
    \},
    \quad
    g_N (q) = \sum_{k=1}^N q_k \, \psi_k(\tau) ,
\end{equation}
the finite-rank approximation (\ref{zJ8aMM}) takes the form
\begin{equation} 
  S_N \epsilon_t 
  = 
  g_N(f_N(\epsilon_t)) ,
\end{equation} 
which is in standard encoder/decoder form. 

We recall that an \emph{encoder/decoder approximation}, also called an \emph{information-based approximation} or \emph{nonlinear reconstruction scheme}, for a mapping $F:X\to Y$ between normed spaces $X$ and $Y$ consists of:
\begin{itemize}
\item[i)] An \emph{encoder} (information operator), $f:X\to \mathbb{R}^N$.
\item[ii)] A \emph{decoder} (reconstruction map), $g:\mathbb{R}^N\to Y$,
\item[iii)] The approximation $F \approx g\circ f$.
\end{itemize}
Thus, the encoder extracts $N$ items of information about the input, while the decoder reconstructs an approximation of the output from this information. 

The encoder/decoder paradigm originates in \emph{information-based complexity} theory \cite{MicchelliRivlin1977, TraubWozniakowski1980, TraubWasilkowskiWozniakowski1988, NovakWozniakowski2008,NovakWozniakowski2010,NovakWozniakowski2012}, which studies the computational cost of approximating operators when only partial information about the input is available. Encoder/decoder approximations generalize classical $N$-width concepts such as the \emph{Kolmogorov width}, or restriction to linear subspaces and projection reconstruction. Encoder/decoder schemes have also been used to describe reduced models and surrogate maps, including reduced basis methods, sparse representations, operator learning and neural operators and model order reduction. \cite{CotterDashtiRobinsonStuart2010, SchwabStuart2012, BhattacharyaHosseiniKovachkiStuart2021, SeidmanKissasPappasPerdikaris2023}
\end{remark}

\subsection{Optimal rank-$N$ approximation of hereditary operator}
\label{subs-optimal}

For a given compact operator $S$ over ${H}$, we wish to identify bases $(\phi_k)_{k=1}^N$, if any, such that the finite-rank approximations (\ref{zJ8aMM}), with $(\psi_k)_{k=1}^N$ then necessarily given by $(S\phi_k)_{k=1}^N$, are {\sl optimal}, in the sense that
\begin{equation} \label{7rPf4y}
  \| S - S_N \| = \inf_{\operatorname{\rm rank} T \leq N} \| S - T \| .
\end{equation}
We note that, in coordinates, this question is equivalent to that of determining the best set of internal variables of a given dimension. It can be shown \cite{Tikhomirov:1960,Pinkus:1985, Ortiz:2025} that the optimality of the operator $S_N$ in the sense (\ref{7rPf4y}) implies an optimal error bound for the corresponding solutions of the Volterra equations (\ref{29a8Zn}) for fixed stress history. 

For Hilbert-Schmidt operators, this problem was first formulated and solved by Schmidt in 1907 \cite{Schmidt:1907} and falls squarely within the theory of $N$-widths \cite{Pinkus:1985}. In the present setting, a central result of the theory is that the optimal rank-$N$ approximation $S_N$ of the history operator $S$ can be characterized in terms of eigenvalues and eigenfunctions of the operators ${S}^*{S}$ and ${S}{S}^*$, where 
\begin{equation}
  S^*\epsilon_t(\tau)
  =
  \int_0^\tau
    \mathbb{C}^{-1}
    \mathbb{K}(\tau-\rho) \, \epsilon_t(\rho)
    \, \frac{w(\rho)}{w(\tau)} 
  \, d\rho ,
\end{equation}
is the {\sl adjoint} anelastic-strain operator in ${H}$. 

Thus, if $S$ is compact, then $S^*S$ and $S S^*$ are compact and self-adjoint operators which define a sequence of positive real eigenvalues $(\mu_k)_{k=1}^{\operatorname{\rm rank}S}$ such that the sequence is non-increasing and, if $\operatorname{\rm rank}S=\infty$, $\lim_{k\to\infty} \mu_k = 0$. With $N \leq \operatorname{\rm rank} S$, let $(\phi_k)_{k=1}^N$ be orthonormal eigenvectors of ${S}^*{S}$ and set $\psi_k = S \phi_k$. Then, $(\psi_k)_{k=1}^N$ are eigenvectors of ${S}{S}^*$ with
\begin{equation}
    \|\psi_k\| = \sqrt{\mu_k(S^*S)} := s_k(S)
\end{equation}
are the $s$-numbers, or {\sl singular values}, of $S$, first introduced by E.~Schmidt \cite{Schmidt:1907} (see also \cite[Chapter IV]{Pinkus:1985}). In addition, 
\begin{equation}
    S_N = \sum_{k=1}^N 
    \psi_k \otimes \phi_k .
\end{equation}
is the best rank-$N$ approximation (\ref{zJ8aMM}) of ${S}$, in the sense of (\ref{7rPf4y}), with (optimal) error estimate
\begin{equation}
  \| S_N - S \| = s_{N+1}(S) ,
\end{equation}
see \cite[Chapter IV]{Pinkus:1985} for a full account. 

\begin{example}[Standard Linear Solid] \label{b5iXcx}
The standard linear solid is characterized by a hereditary operator of the form
\begin{equation} \label{we4WSY}
  (Sf)(\tau)
  =
  \int_{\tau}^{T} k\,{\rm e}^{-\lambda(\rho-\tau)} f(\rho)\,d\rho,
  \quad 
  \tau\in(0,T),
\end{equation}
where $k > 0$ is a relaxation modulus and the operator acts on the space of histories ${H} = L^2\!\left((0,T),\, {\rm e}^{-\lambda_0 \tau}\, d\tau\right)$, $\lambda_0>0$, with inner product 
\begin{equation}
  ( f,g )
  = 
  \int_0^T f(\tau) g(\tau) {\rm e}^{-\lambda_0\tau} \, d\tau.
\end{equation}
A straightforward calculation further gives the adjoint operator as
\begin{equation}
  (S^\ast f)(\tau)
  =
  \int_0^\tau k\,{\rm e}^{-(\lambda-\lambda_0)(\tau-\rho)} f(\rho)\,d\rho.
\end{equation}

\underline{Hilbert-Schmidt property}.
For convenience, we may introduce a unitary operator $U : {H}\to L^2(0,T)$ defined as
\begin{equation}
  (U f)(\tau)
  =
  {\rm e}^{-\frac{\lambda_0}{2}\tau} f(\tau),
\end{equation}
so that
\begin{equation}
  R := U SU^{-1}
\end{equation}
has kernel
\begin{equation}
  K(\tau,\rho) = k\,{\rm e}^{-a(\rho-\tau)} H(\rho-\tau) ,
  \quad 
  a := \lambda+\frac{\lambda_0}{2}.
\end{equation}
Since $R$ is unitarily equivalent to $S$, $S$ is Hilbert--Schmidt iff $R$ is too. Therefore
\begin{equation}
\begin{split}
  \|S\|_{HS}^2 &= \|R\|_{HS}^2 = 
  \int_0^T\int_0^T |K(\tau,\rho)|^2 \,d\rho\,d\tau
  \\ &=
  k^2 \int_0^T\int_\tau^T {\rm e}^{-2 a(\rho-\tau)}\,d\rho\,d\tau 
  =
  k^2 \frac{2 a T+e^{-2 a T}-1}{4 a^2} .
\end{split}
\end{equation}
This bound shows that $S$ is indeed Hilbert--Schmidt, hence bounded, for finite $T$. 

\underline{$N$-width analysis}. Set
\begin{equation}
  \alpha=\lambda-\frac{\lambda_0}{2}.
\end{equation}
Then, $S^*S$ is unitarily equivalent to $A^*A$ on $L^2(0,T)$, where
\begin{equation}
  (Af)(t)=\int_t^T k\, e^{-\alpha(\rho-t)}f(\rho)\,d\rho.
\end{equation}
The operator $A^*A$ is compact, self-adjoint, and satisfies
\begin{equation}
  A^*A=k^2L^{-1},
  \quad
  L=-\frac{d^2}{dt^2}+\alpha^2,
\end{equation}
with boundary conditions
\begin{equation} \label{UFJo4s}
  f(0)=0,
  \quad 
  f'(T) +\alpha f(T)=0.
\end{equation}
Hence the eigenvalue problem 
\begin{equation}
  S^*S\,\xi=\mu\xi ,
\end{equation}
reduces to
\begin{equation} \label{LjG0Ya} 
  -f''(t)+\alpha^2 f(t)=\frac{k^2}{\mu}f(t), 
\end{equation}
subject to (\ref{UFJo4s}). Let $\kappa_n>0$ be the solutions of the transcendental equation
\begin{equation}
  \tan(\kappa_n T) + \kappa_n/\alpha = 0.
\end{equation}
Then, the eigenvalues and normalized eigenfunctions of $S^*S$ are
\begin{equation} \label{5L0uUp}
  \mu_n=\frac{k^2}{\alpha^2+\kappa_n^2},
  \quad
  \phi_n(t) = N_n \, {\rm e}^{\lambda_0 t/2}\sin(\kappa_n t),
  \quad 
  n=1,2,\dots ,
\end{equation}
respectively, with
\begin{equation}
  N_n
  =
  \left(
    \frac{2}{\,T+\dfrac{\alpha}{\alpha^2+\kappa_n^2}}
  \right)^{1/2} .
\end{equation}
In addition, we have
\begin{equation}
  \psi_n(\tau)
  :=
  (S\phi_n)(\tau)
  =
  \frac{k\, N_n}{\alpha^2+\kappa_n^2}\;
  {\rm e}^{\lambda_0\tau/2}
  \Big(\alpha\sin(\kappa_n\tau)+\kappa_n\cos(\kappa_n\tau)\Big) .
\end{equation}

\underline{Special case $\lambda=\lambda_0/2$}. In the special case of $\alpha=0$, the characteristic roots and eigenvalues simplify to:
\begin{equation} \label{d4Etmi}
  \kappa_n=\frac{(n-\tfrac12)\pi}{T},
  \quad
  \mu_n=\frac{k^2T^2}{(n-\tfrac12)^2\pi^2},
  \quad 
  n=1,2,\dots ,
\end{equation}
respectively. In addition, the normalized eigenfunctions reduce to
\begin{equation}
  \phi_n(t)
  =
  \sqrt{\frac{2}{T}}\;e^{\lambda_0 t/2}
  \sin(\kappa_n t) ,
\end{equation}
whence
\begin{equation}
  \psi_n(\tau)
  =
  (S\phi_n)(\tau)
  =
  \frac{k}{\kappa_n}
  \sqrt{\frac{2}{T}}\;e^{\lambda\tau}\cos(\kappa_n\tau).
\end{equation}

\underline{Optimal finite-rank approximation}. Recall that
\begin{equation}
  (S_N\xi)(\tau)
  =
  \sum_{n=1}^N \psi_n(\tau) (\xi,\phi_n) ,
\end{equation}
is the optimal rank-$N$ approximation of $S$.

\begin{figure}
  \centering
  \includegraphics[width=0.95\textwidth]{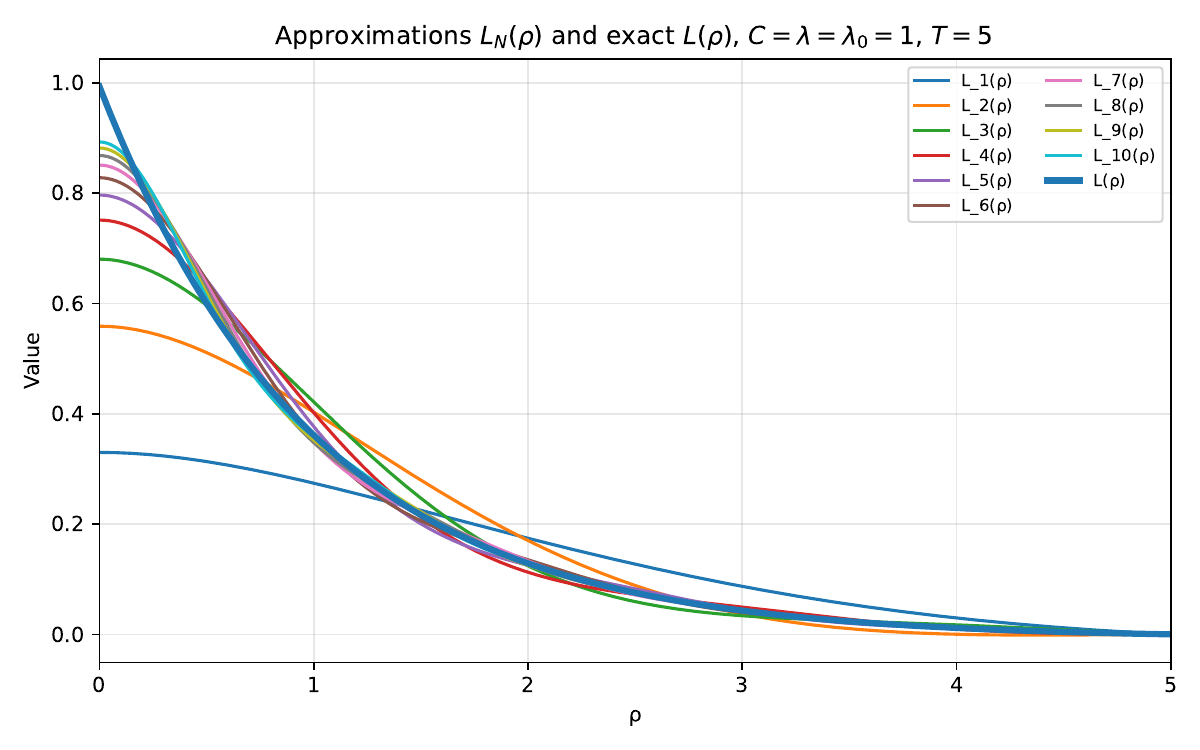}
  \caption{Exact relaxation function and optimal rank-$N$ approximations.} \label{fig:LN_plot}
\end{figure}

The convergence of the sequence $(S_N)$ can be illustrated by testing it with the forward step function $h_\rho(\tau)$ starting at $\rho$. The corresponding relaxation functions
\begin{equation}
  L_N(\rho) := (S_N h_\rho)(0) ,
  \quad
  L(\rho) := (S h_\rho)(0) .
\end{equation}
are shown in Fig.~\ref{fig:LN_plot}. 
The strong convergence of $L_N(\rho)$ to $L(\rho)$ is apparent from the figure. \hfill$\square$
\end{example}

\section{Numerical implementation}

As already stated, we consider viscoelastic systems, such as RVEs, metamaterials or structures, that can be evaluated exactly, i.~e., whose generalized stress histories $\sigma_t(\tau)$ can be evaluated for every generalized strain history $\epsilon_t(\tau)$, albeit at considerable computational expense. For instance, $\epsilon_t(\tau)$ and $\sigma_t(\tau)$ may refer to the average strain and stress of an RVE. However, an explicit analytical representation of the corresponding hereditary law is not available. We wish to determine {\sl optimal} low-rank representations of the unknown hereditary law and address efficient numerical implementations thereof. 

\subsection{Truncation}

We begin by assuming that a weighting function $w(\tau)$ is known that satisfies the conditions of Theorem~\ref{mCpR5D} for all hereditary kernels $\mathbb{K}(\tau)$ of interest, and that a convenient orthonormal basis $(\varphi_i)_{i=1}^\infty$ with respect to the inner product (\ref{KJMHD7}) is available. We proceed to truncate $S$ beyond the $M$th component, with the result
\begin{equation} \label{eNnoDK}
  S_{M,ij}
  :=
  \left\{
    \begin{array}{ll}
      ( \varphi_i, S \varphi_j), & \text{if} \;\, i,j \leq M , \\
      0, & \text{otherwise} .
    \end{array}
  \right.
\end{equation}
Thus, the components $S_{M,ij}$ of the truncated operator are determined by evaluating the histories of inelastic strain $S \varphi_j$ for the basis histories $\varphi_j$, $j=1,\dots,M$, and then computing their first $M$ components in the same basis. 

We recall that we assume that the response of any strain history, in particular histories set forth by the basis functions $\varphi_j$, can be measured or computed exactly. By representing the response $S\varphi_j$ in the same basis, it follows that the evaluation of the components $S_{M,ij}$ of the truncated operator reduces to the evaluation of basis inner products $(\varphi_i, \varphi_j)_H$, which are trivial by the orthonormality of the basis. We also note that $M$ is the number of strain histories for which the material response needs to be \emph{sampled}, i.~e., it represents the \emph{sampling size} from a data perspective.

Finally, we compute the $N$ first eigenfunctions $(\phi_{M,k})_{k=1}^N$ of $S_{M}^T S_{M}$, $N \ll M$, furnishing the optimal rank-$N$ approximation representation of $S_M$ as
\begin{equation} \label{kopnus}
    S_{M,N} = \sum_{k=1}^N \psi_{M,k} \otimes \phi_{M,k} .
\end{equation}
In coordinates, if
\begin{equation}
    \phi_{M,k} = \sum_{i=1}^M \phi_{M,ki}\, \varphi_i , 
    \quad 
    \psi_{M,k} = \sum_{i=1}^M \psi_{M,ki}\, \varphi_i  ,
\end{equation}
then
\begin{equation}
    \psi_{M,ki} = \sum_{j=1}^M S_{M,ij}\, \phi_{M,kj}
\end{equation}
and
\begin{equation} \label{Fg1jFT}
\begin{split} 
    &
    S_{M,N,ij}
    =
    \sum_{k=1}^N \psi_{M,ki}\, \phi_{M,kj} ,
\end{split}
\end{equation}
is the optimal rank-$N$ approximation of $S_M$ in coordinate representation.  

\subsection{Analysis of convergence}

If the operator $S$ is evaluated exactly, then it follows from $N$-width theory, see Appendix~\ref{app:nwidths} and \cite{Tikhomirov:1960,Pinkus:1985}, that the operator norm error incurred by the optimal rank-$N$ approximation is, exactly,
\begin{equation}
  \| S - S_N \| = s_{N+1}(S) .
\end{equation}
If $S$ is a compact operator, see Theorem~\ref{mCpR5D}, then it follows that $\lim_{N\to\infty}$ $s_{N+1}(S)=0$, which establishes the convergence of $S_N$ to $S$ in the operator norm.

However, in the procedure outlined in the foregoing, $S$ itself is approximated by truncation, which introduces additional errors to be estimated. Thus, triangulating,
\begin{equation} \label{n6lmWQ}
  \| S - S_{M,N} \|
  \leq
  \| S - S_{M} \| + \| S_{M} - S_{M,N} \| .
\end{equation}
In addition, an appeal to duality yields
\begin{equation}
\begin{split}
  s_{M,N+1}(S)
  &:=
  \| S_{M} - S_{M,N} \|
  =
  \inf_{\operatorname{\rm dim} V \leq N}
  \sup_{(\eta,V)=0}
  \frac{\|S_{M}^*\eta\|}{\|\eta\|} \\
  &\leq
  \inf_{\operatorname{\rm dim} V \leq N}
  \Big(
    \sup_{(\eta,V)=0}
    \frac{\|S^*\eta\|}{\|\eta\|}
    +
    \sup_{(\eta,V)=0}
    \frac{\|(S-S_{M})^*\eta\|}{\|\eta\|}
  \Big) \\
  &\leq
  s_{N+1}(S)
  +
  \sup_{(\eta,V_S)=0}
  \frac{\|(S-S_{M})^*\eta\|}{\|\eta\|}
  \\
  &\leq
  s_{N+1}(S)
  +
  \|S-S_{M}\| ,
\end{split}
\end{equation}
where $V_S$ is optimal with respect to $S$. Inserting this in (\ref{n6lmWQ}),
\begin{equation} \label{NrO1E0}
  \| S - S_{M,N} \|
  \leq
  s_{N+1}(S)
  +
  2 \| S - S_{M} \| ,
\end{equation}
which shows that the additional error is controlled by $\| S - S_{M} \|$. Let
\begin{equation}
  \Pi_M \xi = \sum_{i=1}^M (\varphi_i,\xi) \varphi_i
\end{equation}
be the orthogonal projection of ${H}$ onto $\operatorname{\rm span}(\varphi_1,\dots,\varphi_M)$. Then, the truncated operator (\ref{eNnoDK}) follows as
\begin{equation}
  S_M = \Pi_M S \Pi_M .
\end{equation}
The orthogonal projections satisfy $\Pi_M \xi \to \xi$ for every $\xi\in H$,
i.e., $\Pi_M\to I$ strongly on $H$. A standard result \cite{Conway:1990} states that if
$\Pi_M \to I$ strongly and $S$ is compact, then
\begin{equation}
\|(I-\Pi_M)S\| \to 0,
\qquad \text{and} \qquad
\|S(I-\Pi_M)\| \to 0,
\end{equation}
as $M\to\infty$. Now observe that
\begin{equation}
S-\Pi_M S\Pi_M
=
(I-\Pi_M)S+\Pi_M S(I-\Pi_M).
\end{equation}
Therefore,
\begin{equation}
\|S-\Pi_M S\Pi_M\|
\le
\|(I-\Pi_M)S\|+\|\Pi_M\|\,\|S(I-\Pi_M)\|.
\end{equation}
Since $\Pi_M$ is an orthogonal projection, $\|\Pi_M\|=1$, and thus
\begin{equation}
\|S-\Pi_M S\Pi_M\|
\le
\|(I-\Pi_M)S\|+\|S(I-\Pi_M)\|
\to 0,
\end{equation}
as $M\to\infty$. Hence,
\begin{equation}
\|S-S_M\|=\|S-\Pi_M S\Pi_M\|\to 0,
\end{equation}
which is the required convergence of the truncated operators in operator norm. Consequently, from
\begin{equation}
\|S-S_{M,N}\|
\le
\|S-S_M\|+\|S_M-S_{M,N}\|
=
\|S-S_M\|+s_{N+1}(S_M),
\end{equation}
and the estimate
\begin{equation}
s_{N+1}(S_M)\le s_{N+1}(S)+\|S-S_M\|,
\end{equation}
it follows that
\begin{equation}
\|S-S_{M,N}\|
\le
s_{N+1}(S)+2\|S-S_M\|,
\end{equation}
with $\|S-S_M\|\to 0$ as $M\to\infty$. Since $S$ is compact,
$s_{N+1}(S)\to 0$ as $N\to\infty$, and therefore
\begin{equation}
\|S-S_{M,N}\|\to 0,
\end{equation}
whenever $M\to\infty$ and $N\to\infty$.

Collecting the preceding estimates, we finally arrive at the error bound
\begin{equation} \label{wS2ouV}
  \| S - S_{M,N} \|
  \leq
  s_{N+1}(S)
  +
  2\bigl(\|(I-\Pi_M)S\|+\|S(I-\Pi_M)\|\bigr) .
\end{equation}
This establishes the convergence of the scheme. The second term in this bound may be interpreted as a \emph{sampling error} and the first term as a \emph{rank error}, the total error being bounded by the sum of both.

We note, however, that the analysis, as it stands, does not supply a rate of convergence, which requires precise quantitative estimates of the decay of
$\|(I-\Pi_M)S\|$, $\|S(I-\Pi_M)\|$, and $s_{N+1}(S)$, together with detailed consideration of regularity properties.

\section{Numerical tests} \label{sec-history}

In this section, we present two numerical examples that illustrate the approximation properties of the optimal history representation presented in the foregoing: i) a simple one-dimensional standard linear solid; and ii) a representative volume element (RVE) in the form of a regular polycrystal, intended to exemplify how the method of approximation applies to situations where the material behavior is characterized by way of multiscale analysis. It bears emphasis that similar procedures apply \emph{mutatis mutandi} when material behavior is characterized experimentally. 

\subsection{One-dimensional example} \label{subs-1d}

\begin{figure}[p]
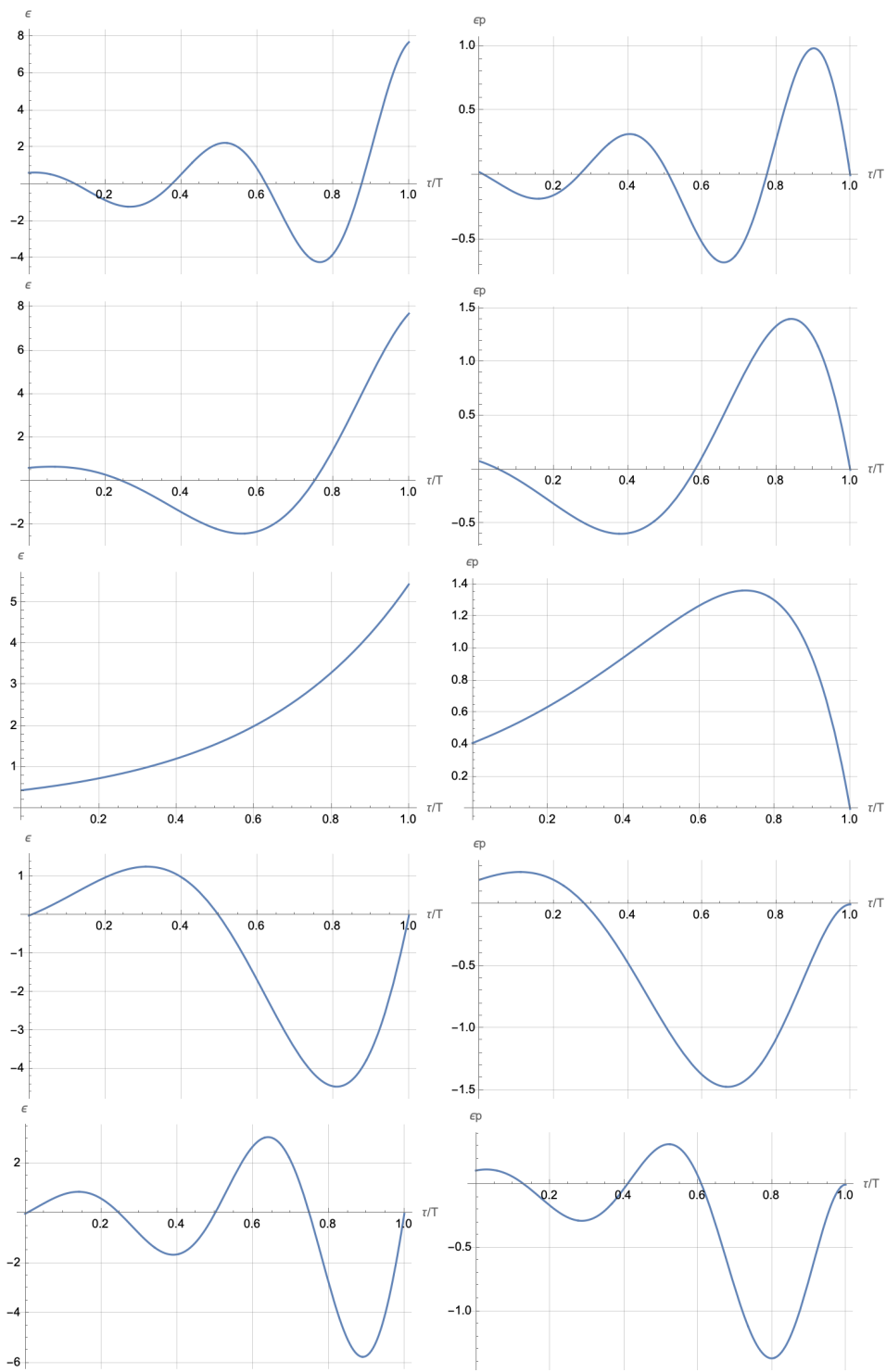

    \centering
    \foreach \i in {-2,...,2}
    {
        \includegraphics[width=0.49\textwidth]{OD-eh-\i.png}
        \includegraphics[width=0.49\textwidth]{OD-eph-\i.png}
    }
    \caption{Standard linear solid, (\ref{eq-ex1-S}), $C_0=2$, $C_1=1$, $\lambda=\lambda_0=1$. From
      top to bottom, rows show a basis function $e_n(\tau)$ (left) and the corresponding
      viscoelastic response $e_n^{p}(\tau)\equiv S e_n(\tau)$ (right),
      with $n=-2,\ldots,2$.} \label{fig-basis}
\end{figure}

We consider a one-dimensional standard linear solid characterized by a hereditary law (\ref{29a8Zn}) of the form
\begin{equation} \label{eq-ex1-sigma}
    \sigma(t) 
    = 
    C_0 \epsilon(t) 
    - 
    \int_0^t C_1 e^{-\lambda (t-s)} \epsilon(s) \, ds .
\end{equation}
where $C_0 > 0$, $C_1 > 0$ and $\lambda > 0$ are material constants, cf.~Example~\ref{b5iXcx}, and the strain and stress histories are assumed to vanish for $t < 0$. Alternatively, in the history representation (\ref{ZrANem}) 
\begin{equation} \label{eq-ex1-S}
    \epsilon^p_t(\tau) 
    = 
    \frac{\sigma_t(\tau)}{C_0} - \epsilon_t(\tau) 
    =
    \int_{\tau}^T \frac{C_1}{C_0} \, {\rm e}^{-\lambda(\rho-\tau)} \, \epsilon_t(\rho) \,d\rho
    :=  
    S \epsilon_t(\tau) .
\end{equation}
where $S$ is the inelastic-strain operator, cf.~(\ref{we4WSY}), and we assume that the strain and stress histories have finite duration $T$. We further assume that all histories belong to the Hilbert space $H \equiv L^2((0,T), {\rm e}^{-\lambda_0\tau}\,d\tau)$, with $\lambda_0>0$ satisfying condition (\ref{DcjG6H}). A convenient basis for representing histories in $H$ is
\begin{equation} \label{eq-ex1-basis}
    e_n(\tau)
    =
    \begin{cases}
        \sqrt{\frac{2}{T}} \, {\rm e}^{\lambda_0 \tau/2}\cos(\frac{2\pi n\tau}{T}) ,
        & n < 0 , \\ 
        \sqrt{\frac{1}{T}} \, {\rm e}^{\lambda_0 \tau/2} ,
        & n=0 , \\
        \sqrt{\frac{2}{T}} \, {\rm e}^{\lambda_0 \tau/2}\sin(\frac{2\pi n\tau}{T}) ,
        & n >0 . 
    \end{cases}
\end{equation}
Fig.~\ref{fig-basis} shows these functions for $-2 \le n \leq 2$, together with the
corresponding viscoelastic response in the standard linear model. 

Next, we turn to the optimal approximation of the viscoelastic operator~$S$, see
Section~\ref{subs-optimal}. For that, we select a finite basis of histories $(e_n(\tau))_{n=-m}^m$
of size $M=2m+1$. Then, in coordinates, the inelastic-strain operator takes the form
\begin{equation} \label{eq-SMij}
    S_{M,ij} 
    = (e_{\Sigma(i)},Se_{\Sigma(j)})_H ,
    \quad 
    1 \le i,j \le M ,
\end{equation}
where we introduce the index shift map $\Sigma(i)=i-m-1$ and calculate $Se_{\Sigma(j)}$ using Eq.~\eqref{eq-ex1-S}. Given the simple form of the basis functions, the components of $S_M$ can be calculated analytically in closed form. Next, for $N\ll M$ we calculate the first $N$ eigenvalues of $S^T_MS_M$, together with the corresponding normalized eigenvectors $\phi_{M,k},\; k=1,\ldots,N$ and functions $\psi_{M,k}=S_M\phi_k$. Then, according to $N$-width theory, the best rank-$N$ approximation $S_{M,N}$ of $S_M$ in the $M$-dimensional space spanned by $(e_n)_{n=-m}^{m}$ is given by (\ref{kopnus}). 

\begin{figure}[p]
    \centering
    \foreach \i in {1,...,10}
    {
        \includegraphics[width=0.49\textwidth]{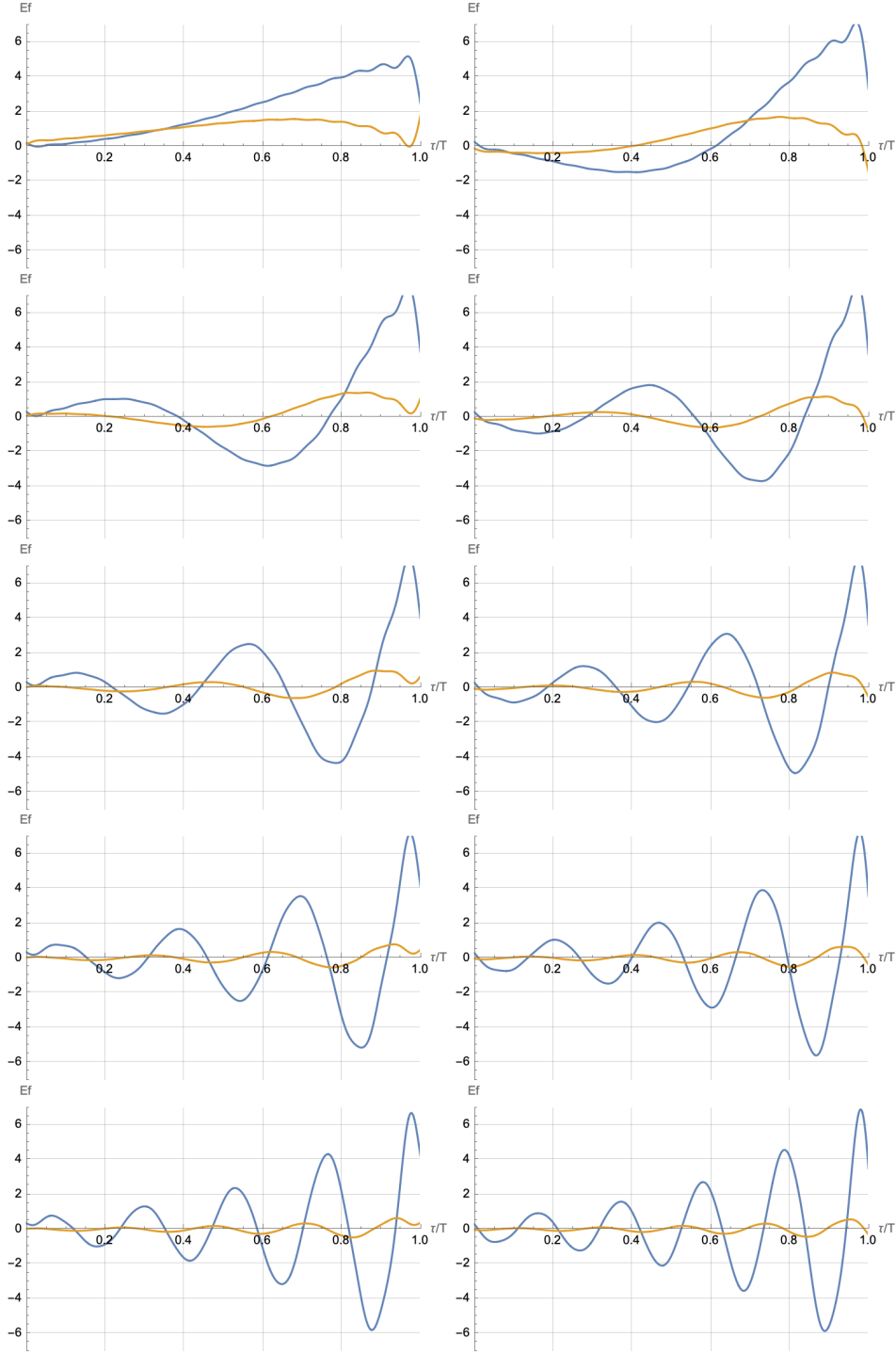}%
    }
    \caption{Standard linear solid, (\ref{eq-ex1-S}), $C_0=2$, $C_1=1$, $\lambda=\lambda_0=1$, $m=15$, $M=2m+1 = 31$. Right eigenfunctions $\phi_{M,k}$ (blue) and left eigenfunctions $\psi_{M,k}$ (orange). From top to bottom, left to right, $k=1,\dots,10$.} \label{fig-ex1-ef}
\end{figure}

\begin{figure}[ht]
    \centering 
    \includegraphics[width=0.75\textwidth]{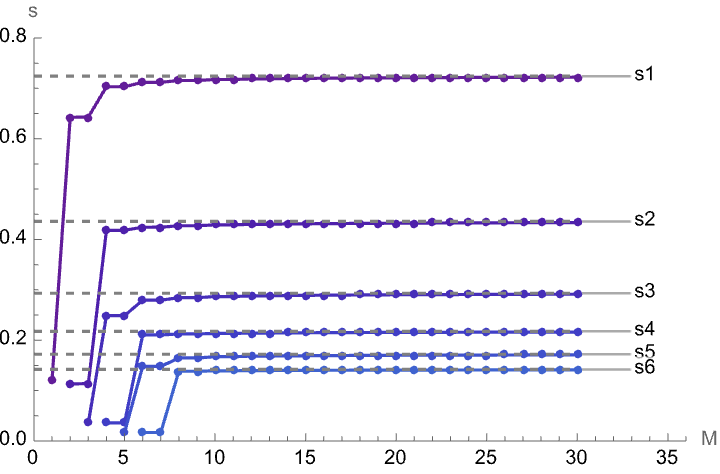} \caption{Standard linear solid, (\ref{eq-ex1-S}), $C_0=2,\ C_1=1, \lambda=\lambda_0=1$. Singular values $s_{M,k}$, $k=1,2,\ldots,6$ of $S_{M}$ as a function of $M$, the dimension of the truncated history subspace. Horizontal lines depict exact analytical values, $M=\infty$.} \label{fig-ex1-eval}
\end{figure}

Fig.~\ref{fig-ex1-ef} depicts the right and left eigenfunctions, $\phi_{M,k}$ and $\psi_{M,k}$,
respectively, of the truncated viscoelastic operator $S_M$, for $m=15$, $M=2m+1 = 31$.
Fig.~\ref{fig-ex1-eval} shows the singular numbers $(s_{M,k})_{k=1}^6$ of $S_M$ as a function of
$M$. The exact analytical values $s_k$, $M=\infty$, are also shown for comparison. The convergence
of $s_{M,k}$ to $s_k$ as $M\to\infty$ is evident in the figure and bears out~(\ref{NrO1E0}).

\begin{figure}[p] 
    \centering
    {
        \includegraphics[width=0.49\textwidth]{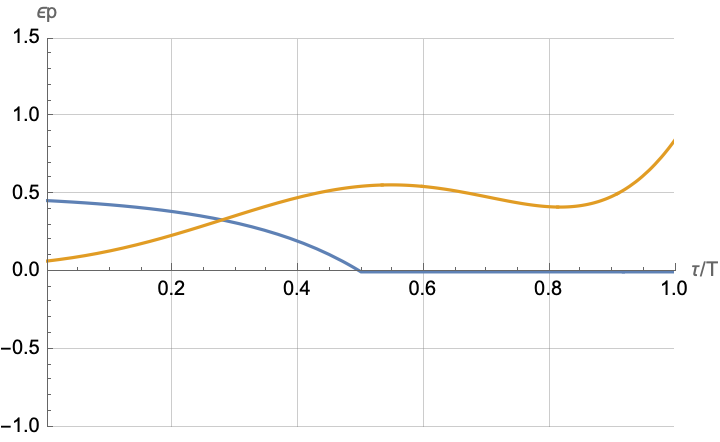}%
        \includegraphics[width=0.49\textwidth]{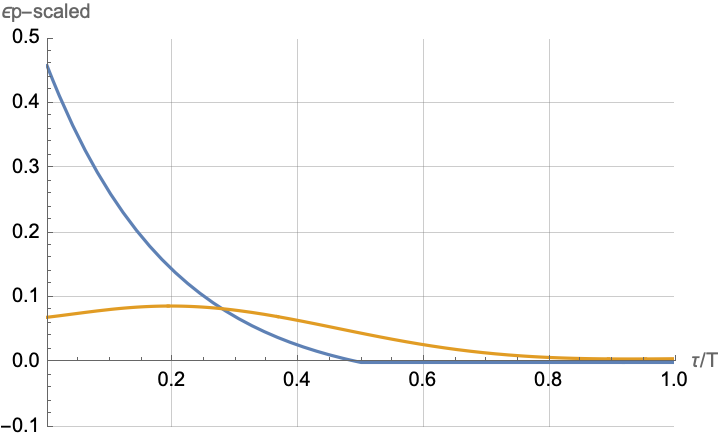}
        \includegraphics[width=0.49\textwidth]{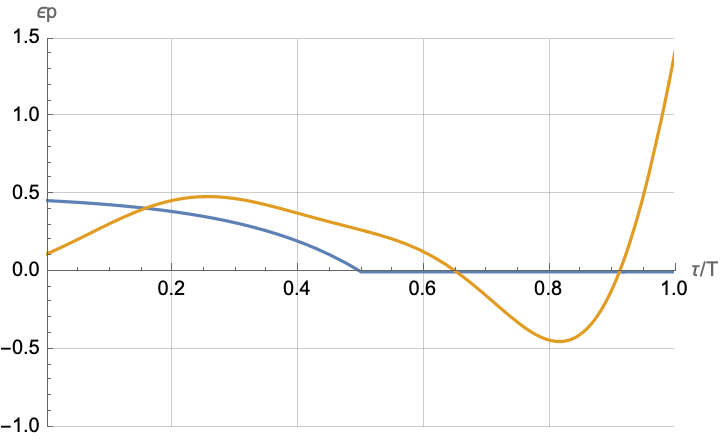}%
        \includegraphics[width=0.49\textwidth]{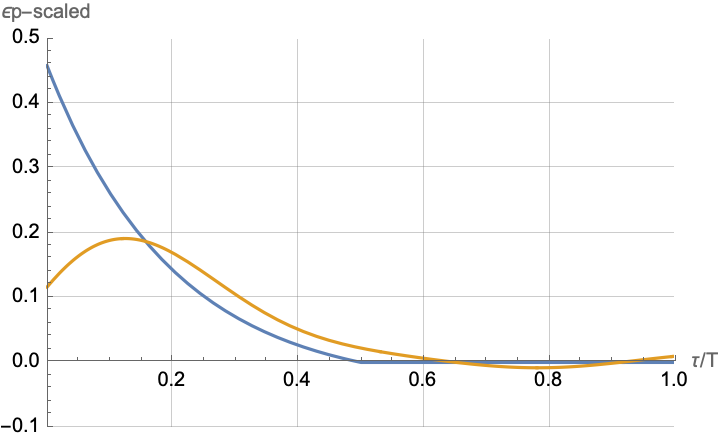}
        \includegraphics[width=0.49\textwidth]{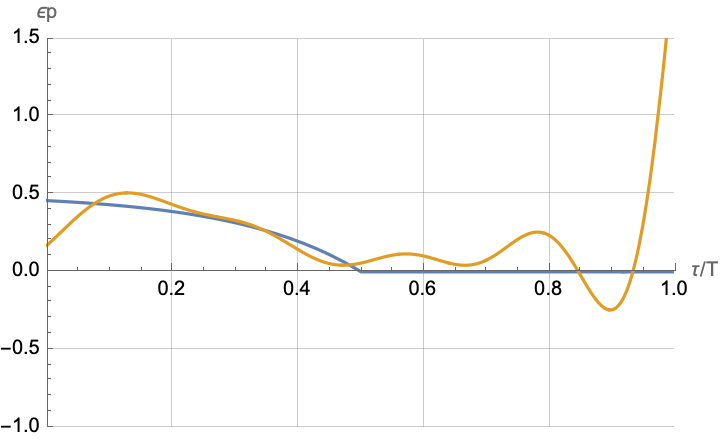}%
        \includegraphics[width=0.49\textwidth]{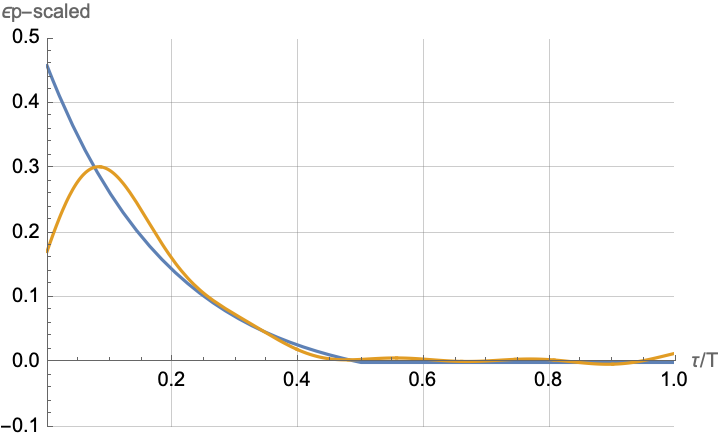}
        \includegraphics[width=0.49\textwidth]{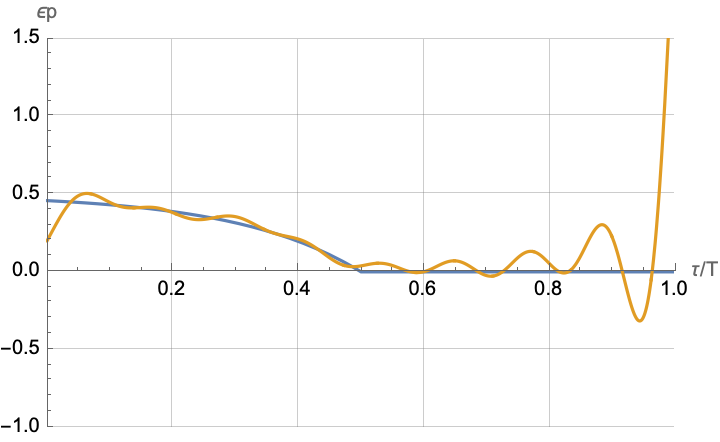}%
        \includegraphics[width=0.49\textwidth]{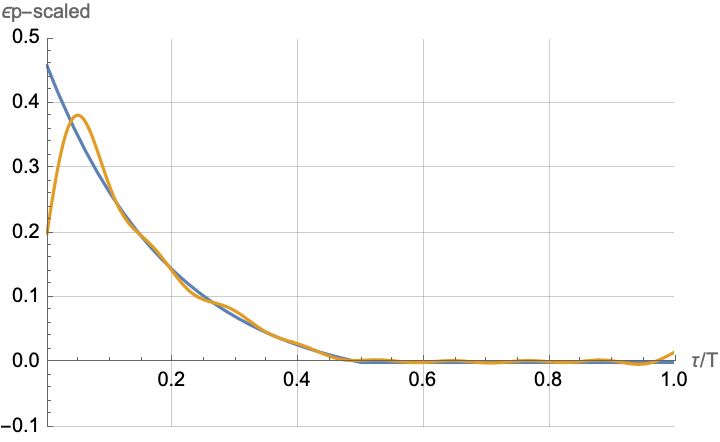}
        \includegraphics[width=0.49\textwidth]{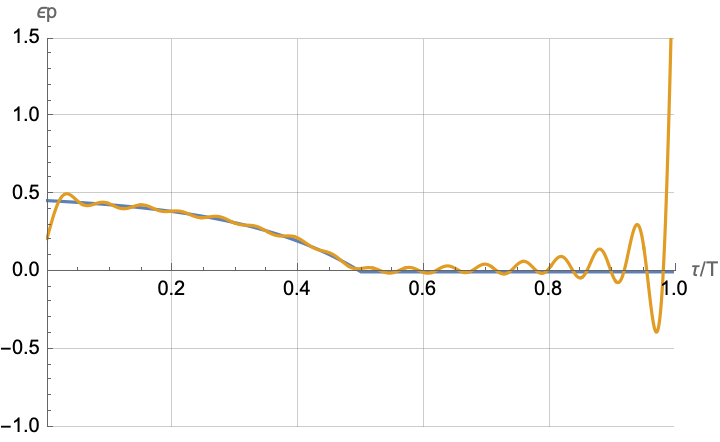}%
        \includegraphics[width=0.49\textwidth]{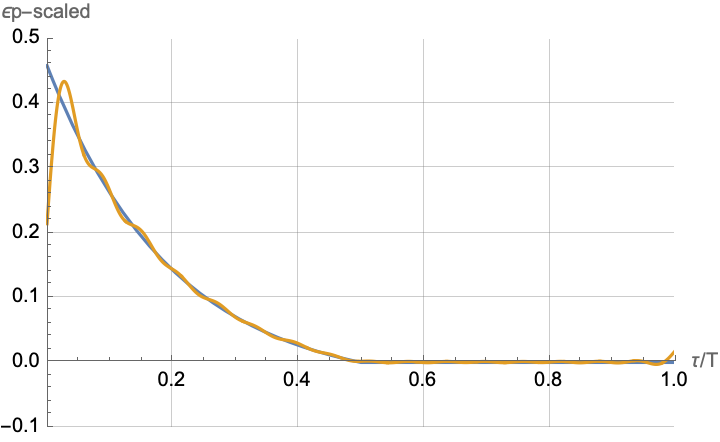}
    }
    \caption{Standard linear solid, (\ref{eq-ex1-S}), $C_0=2$, $C_1=1$, $\lambda=\lambda_0=1$. In orange, inelastic strain (left) and weighted inelastic strain (right) when the model is loaded with a unit step strain at $\tau/T=0.5$. From top to bottom, solutions obtained with optimal history representations of dimension $N=1,2,4,8,16$ and basis of dimension $M=2N+1$. The exact solution is shown in blue.
    } \label{fig-ex1-step}
\end{figure}

\begin{figure}[ht]
    \centering
    \includegraphics[width=0.7\textwidth]{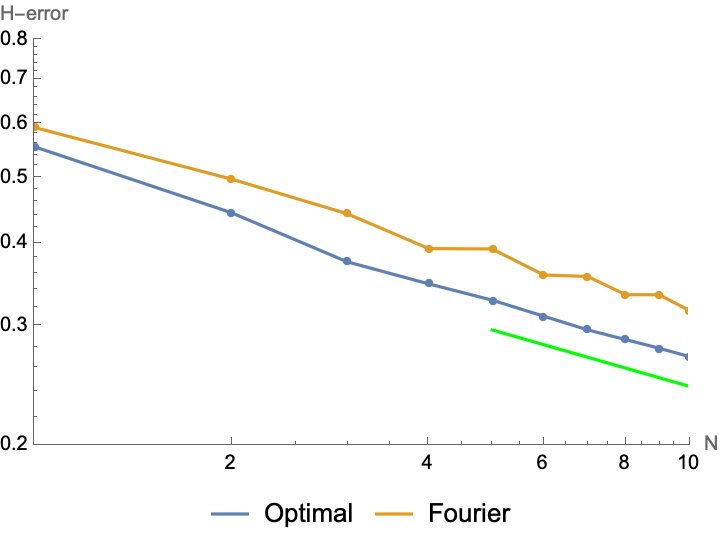}
    \caption{Standard linear solid, (\ref{eq-ex1-S}), $C_0=2$, $C_1=1$, $\lambda=\lambda_0=1$.
      Error in the inelastic strain measured in the $H$-norm as a function
      of $N$, the size of the eigen-basis, with a representation space of dimension $M=2N+1$.
      Optimal \emph{vs.}~suboptimal Fourier-type representation. A reference curve
      with slope $-0.28$ is shown in green.}
    \label{fig-step-error}
\end{figure}

Next, we examine the convergence of the optimal finite-rank approximations of the viscoelastic operator and
verify their optimality. To that end, we consider the step strain history
\begin{equation} \label{eq-ex1-step}
    \epsilon_t(\tau) 
    =
    \begin{cases}
        1, & 0\le\tau\le T/2 , \\
        0, & T/2 < \tau \le T .
    \end{cases}
\end{equation}
The inelastic strain history is
\begin{equation} \label{eq-ex1-step-ep}
    \epsilon^p_t(\tau) 
    =
    \begin{cases}
        \frac{C_1}{C_0\,\lambda}
        \left(
            1 - {\rm e}^{-\lambda(T/2-\tau)}
        \right) , 
        & 0 \le \tau \le T/2 , \\
        0 , & T/2 < \tau < T ,
    \end{cases}
\end{equation}
and the corresponding stress history is
\begin{equation} \label{eq-ex1-step-s}
    \sigma_t(\tau) 
    =
    \begin{cases}
        C_{0} + \frac{C_1}{\lambda}
        \left(
            {\rm e}^{-\lambda(T/2-\tau)} - 1
        \right) ,
        & 0 \le \tau \le T/2 , \\ 
        0, & T/2 < \tau < T .
    \end{cases}
  \end{equation}
  
From these closed-form expressions, we may obtain the error, measured in the $H-$norm, of the inelastic strain history
$S_{M,N} \epsilon_t$, Eqs.~(\ref{kopnus}) or (\ref{Fg1jFT}), as a function of $N$, with $M=2N+1$.
Fig.~\ref{fig-step-error} shows this error as $N$ increases. 
For purposes of comparison, we also show the approximation error when a sub-optimal basis of the same size $N$ is
selected. Specifically we use the scaled Fourier basis~\eqref{eq-ex1-basis}. As expected from (\ref{n6lmWQ}), the error decreases with increasing $N$ down to a floor value due to truncation. Also as expected, the error incurred by the optimal approximation is significantly less than the error resulting from the suboptimal representation.

A striking feature of the approximating histories in Fig.~\ref{fig-ex1-step} is the Gibbs phenomenon that is evident in the distant past, as $\tau$ approaches $T$. For the standard-linear-solid test used in the example, the exact inelastic strain produced by a step strain history is
\begin{equation}
    \epsilon_t^p(\tau)
    =
    \begin{cases}
        \dfrac{C_1}{C_0\lambda}\Big(1-e^{-\lambda(T/2-\tau)}\Big), 
        & 0\le \tau\le T/2,\\[1ex] 
        0, & T/2<\tau<T,
    \end{cases}
\end{equation}
so the target history is only piecewise smooth and has a sharp cutoff. A finite sum of smooth global oscillatory modes cannot reproduce such localized nonsmooth behavior pointwise and, instead, it distributes the error into oscillatory ringing patterns. This is the same mechanism as in classical Fourier truncation, see, e.g.,  \cite{GottliebShu1997Gibbs}. Importantly, the oscillatory approximation does converge properly in the intended weighted $L^2$-norm of $H$, but it does signal a certain lack of control in a stronger sense (for connections with the fading memory phenomenon, see the seminal paper of G.~Fichera \cite{Fichera1979MemoriaTenace}).

More precisely, we note that the singular functions used in the optimal rank-$N$ approximation come from the Sturm--Liouville problem (\ref{UFJo4s}) and (\ref{LjG0Ya}),  with eigenvalues and normalized eigenfunctions (\ref{5L0uUp}) and, in the special case $\alpha=0$, wavenumbers $\kappa_n$ as in (\ref{d4Etmi}). Hence, the retained modes oscillate up to a largest wavenumber
\begin{equation} \label{OBjHSt}
    \kappa_{\max}\sim \kappa_N \sim \frac{\pi N}{T},
\end{equation}
but do \emph{not} impose the exact terminal behavior of the target history at $\tau=T$. Instead, they only satisfy the mixed boundary condition of the second identity in Eq.~(\ref{UFJo4s}). Consequently, the reconstruction of a history with a sharp cutoff from finitely many such modes necessarily results in an endpoint \emph{boundary layer} with oscillations concentrated near $\tau=T$.  

The characteristic width of the Gibbs-type boundary layer near $\tau = T$ can be estimated simply as follows. Since the truncated history space uses modes with largest resolved wavenumber (\ref{OBjHSt}), the boundary-layer width is set by the inverse smallest resolved length scale, i.e., 
\begin{equation}
    \delta_{\mathrm{BL}} \asymp \frac{1}{\kappa_{\max}}.
\end{equation}
If the finite-rank approximation error dominates, we then have $\delta_{\mathrm{BL}} \sim \frac{T}{N}$ for the rank-$N$ optimal approximation. Contrariwise, if the effect is controlled by the $M$-mode truncation of the history basis, then $\delta_{\mathrm{BL}} \sim \frac{T}{M}$. Therefore, for general approximation,
\begin{equation}
    \delta_{\mathrm{BL}} \sim \max\!\left(\frac{T}{M},\,\frac{T}{N}\right).
\end{equation}
For the choice, $M = 2N+1$, this estimate becomes
\begin{equation}
    \delta_{\mathrm{BL}} \sim \frac{T}{N} \sim \frac{2T}{M},
\end{equation}
up to a constant factor of order $1$ depending on how the width is defined
(first zero, first extremum, visible oscillation envelope, etc.).

In practice, the preceding analysis suggests a number of strategies for eliminating the Gibbs effect, if so desired. One option is to discard the solution after a time of the order $\delta_{\mathrm{BL}}$ prior to $T$. Alternatively, the applied strain history may be extended by $0$ up to a time exceeding $T$ by an interval of order $\delta_{\mathrm{BL}}$. A more ambitious strategy is to seek approximations controlled by a stronger norm than the weighted $L^2$-norm considered in this work, but such extensions are beyond the scope of the paper. 

\subsection{Viscoelastic response of an idealized polycrystal} \label{subs-poly}
\begin{figure}[t]
  \centering
  \includegraphics[width=0.9\textwidth]{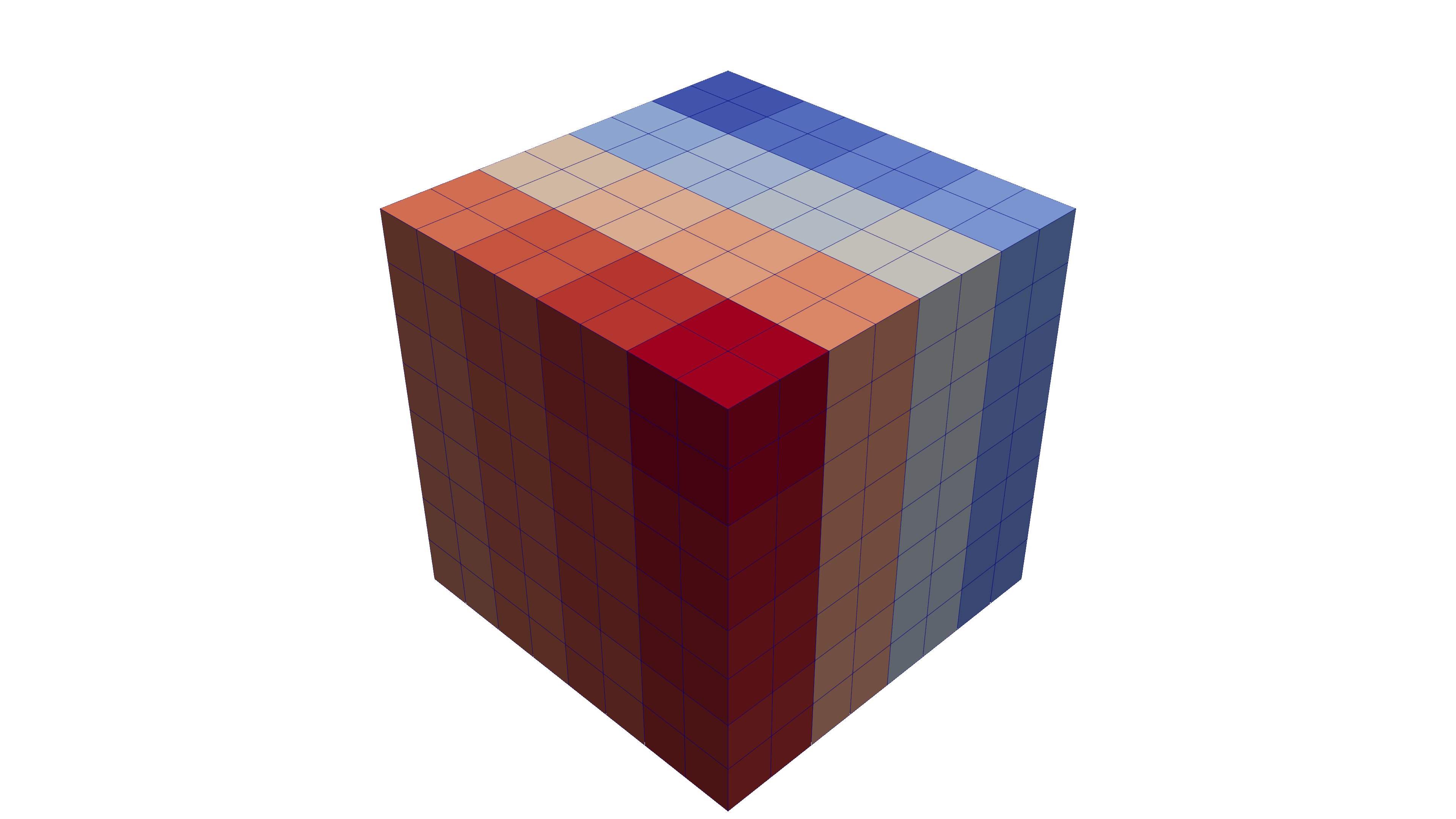}
  \caption{Representative volume element and its finite element mesh. The RVE has $4^3$ ``grains'', each of them with a different viscoelastic behavior and assigned to a different color. The finite element mesh employs 8 hexahedral elements per grain.} \label{fig-RVE}
\end{figure}

\begin{figure}[ht]
  \centering
  \includegraphics[width=0.49\textwidth]{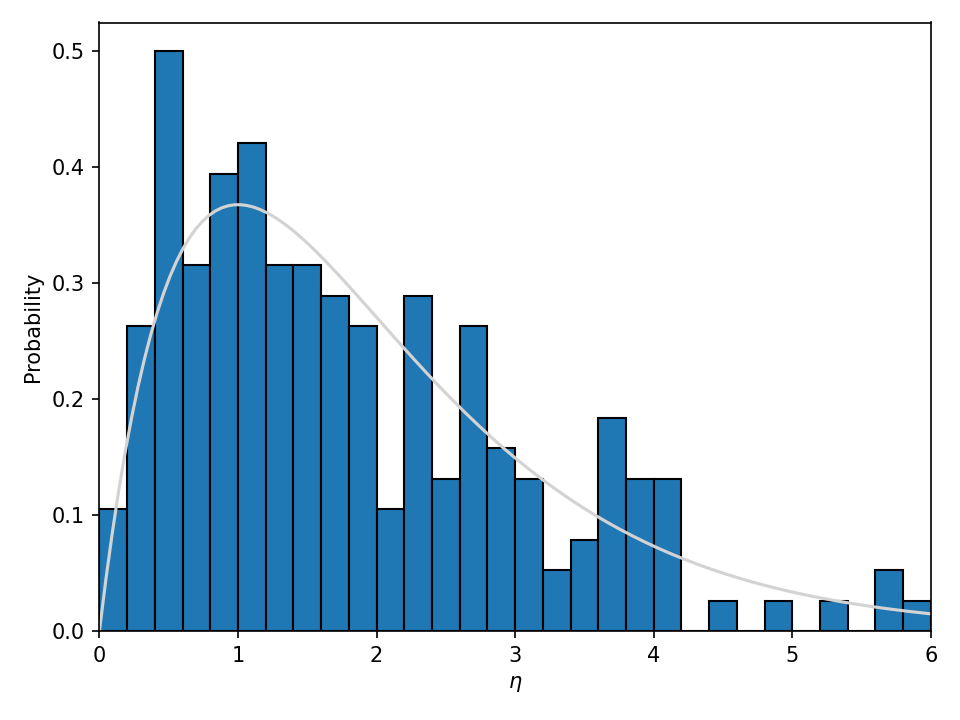}%
  \includegraphics[width=0.49\textwidth]{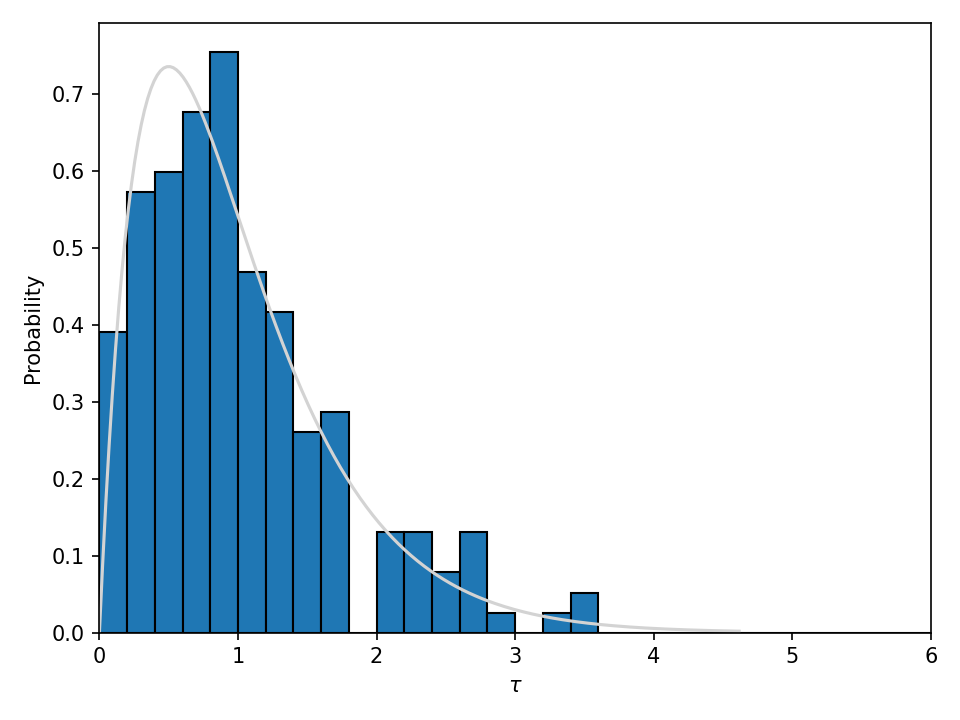}
  \caption{RVE example. Histograms with the statistical distribution of the viscosity $\eta_{ik}$ and characteristic times $\tau_{ik}$ of all the materials in the RVE. The Gamma distributions where the two random variables are sampled from are superposed on top of the data.} \label{fig-etatau}
\end{figure}

Next, we examine the ability of the theory to represent the viscoelastic response of a complex material. To that end, we consider a periodic, inhomogeneous, representative volume element (RVE) in the form of a cube of unit volume consisting of $N_g=4^3$ cubic regions, or \emph{grains}, each of them viscoelastic and isotropic (see Fig.~\ref{fig-RVE} for an illustration of this RVE). The volumetric response of the grains is elastic and homogeneous, with bulk modulus $\kappa=5/3$. By contrast, the deviatoric response varies from grain to grain and obeys a Wiechert model with $W=3$ Maxwell elements and a uniform long-term shear modulus $\mu_{\infty}=1$, see Example~\ref{t3QNXc}. The viscosity of each Maxwell element is sampled from a Gamma distribution of mean $\bar{\mu}=2$ and shape $\bar{\kappa}=2$. Similarly, the characteristic time of each Maxwell element is sampled from another Gamma distribution with mean $\bar{\tau}=1$ and shape $\bar{\beta}=2$. Fig.~\ref{fig-etatau} shows histograms of the two sampled random variables. Under these assumptions, the relaxation modulus of grain $k$ is 
\begin{equation} \label{eq-ex2-relaxation}
    \mathbb{R}_k(\tau) 
    = 
    \mu_{\infty} + \sum_{i=1}^W \mu_{ki}\; {\rm e}^{-\tau/\tau_{ki}}\ ,
    \quad
    \mu_{ki}=\frac{\eta_{ki}}{\tau_{ki}},
    \quad
    k=1,\dots, N_g .
\end{equation}
where $\eta_{ki}$ and $\tau_{ki}$ are the viscosities and characteristic times of grain $k$, respectively. 

\begin{figure}[p]
    \centering
    \foreach \i in {1,...,2}{
      \includegraphics[width=0.4\linewidth]{PC-strain-\i.png}
      \includegraphics[width=0.4\linewidth]{PC-stress-\i.png}\\}
    \includegraphics[width=0.4\linewidth]{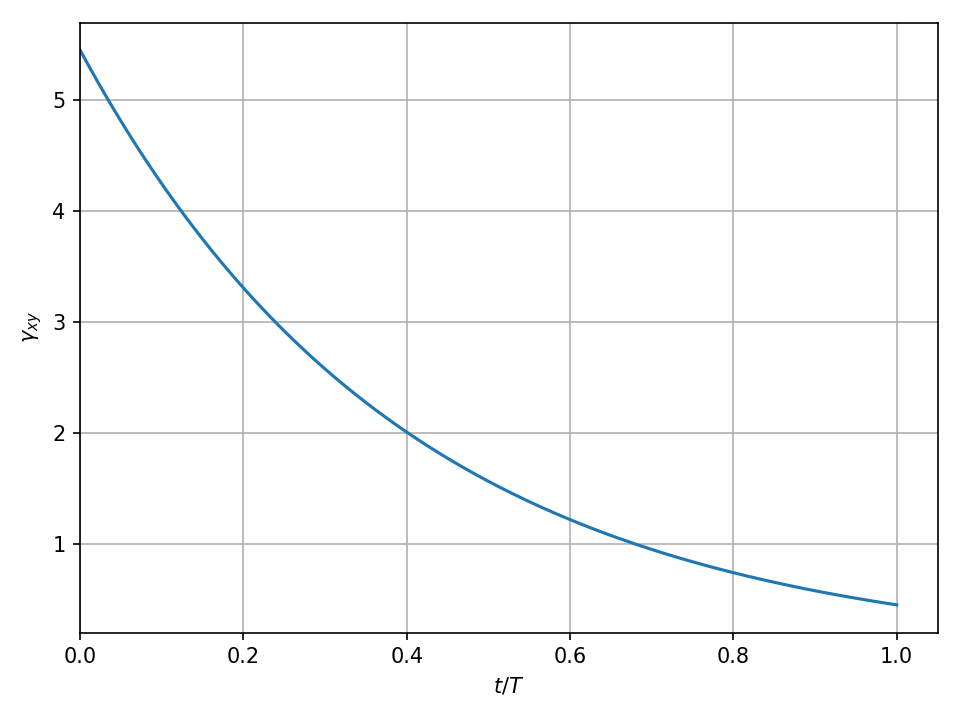}
    \includegraphics[width=0.4\linewidth]{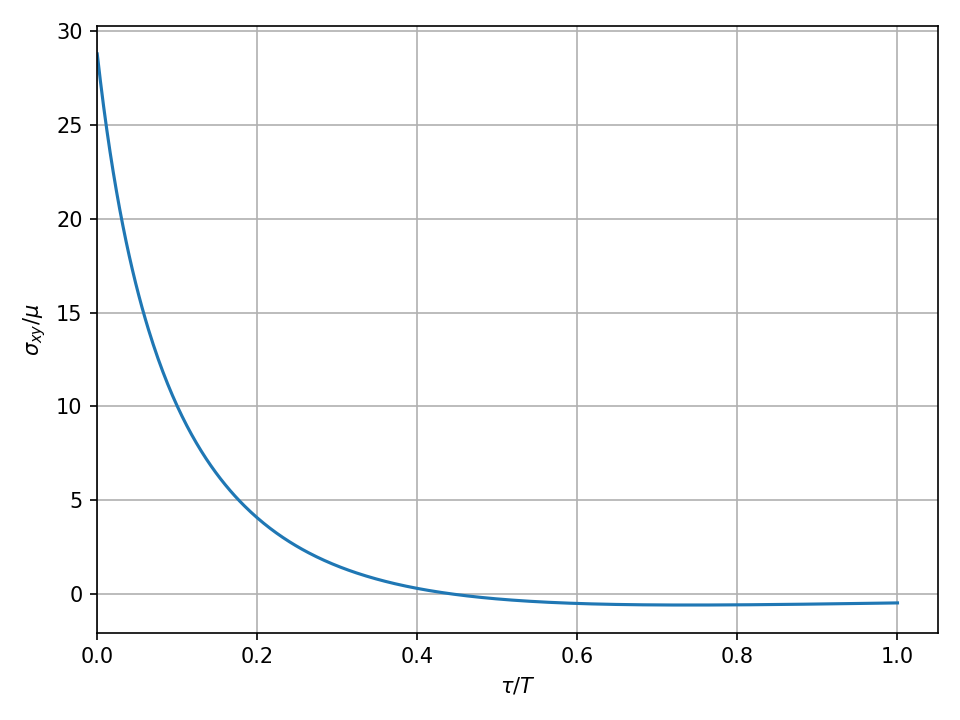}\\
    \foreach \i in {21,...,22}{%
      \includegraphics[width=0.4\linewidth]{PC-strain-\i.png}
      \includegraphics[width=0.4\linewidth]{PC-stress-\i.png}\\}
    \caption{RVE example. Imposed strains $\epsilon_{xy}(t)=e_j(T-t)$ (left) vs. normalized mean
      stress $\tau_j(t)/\mu_{\infty}$ (right), $-2 \le j \le 2$ computed
      with the FE model.}
    \label{fig-poly-strainstress}
\end{figure}

The effective, or \emph{homogenized}, response of the RVE follows as in the derivation of Section~\ref{NOtUv7}, to which we append periodic boundary conditions. We specifically evaluate the RVE by means of a finite element mesh consisting of $8^3$ hexahedral elements, Fig.~\ref{fig-RVE}, with prescribed macroscopic shear strain histories of period $T=5$ represented by means of basis \eqref{eq-ex1-basis}. We select $m=20$ in this basis and calculate, using the finite element discretization of the RVE, the response to $M=2m+1=41$ prescribed macroscopic shear strain histories. 

\begin{figure}[p]
    \centering
    \foreach \i in {1,...,8} 
    {
      \includegraphics[width=0.49\linewidth]{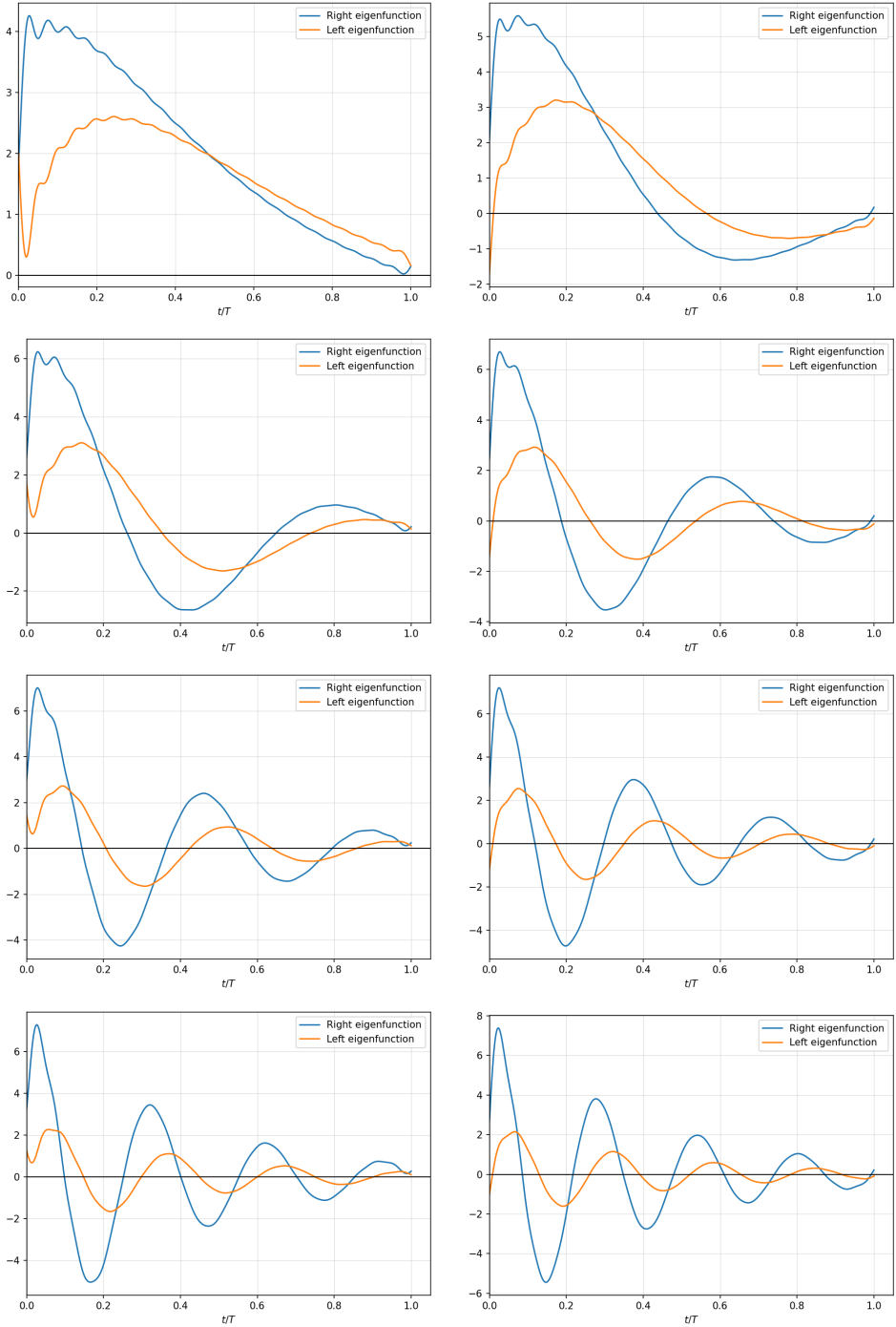}
    }
    \caption{RVE example. Right and left eigenfunctions of the operator $S_M$. From left to right,
      top to bottom, eigenfunctions $\phi_k,\psi_k$ with $k=1,\ldots,8$, and $M=41$.} \label{fig-poly-eigenfunctions}
\end{figure}

Fig.~\ref{fig-poly-strainstress} shows the prescribed shear strains $e_j(T-t)$ and the corresponding shear stresses $\tau_j(t)$, for $-2\le j\le 2$. Following the same steps as in Section~\ref{subs-1d}, from these data, we calculate the truncated inelastic-strain operator $S_M$, its right and left eigenfunctions, and its singular values. Fig.~\ref{fig-poly-eigenfunctions} shows the first $8$ right and left eigenfunctions of the $S_M$ as functions of time. Finally, the first $N\ll M$ eigenfunctions of $S_M$ then determine its optimal rank-$N$ approximation $S_{M,N}$. 
 
\begin{figure}[ht]
  \centering
  \includegraphics[width=0.49\textwidth]{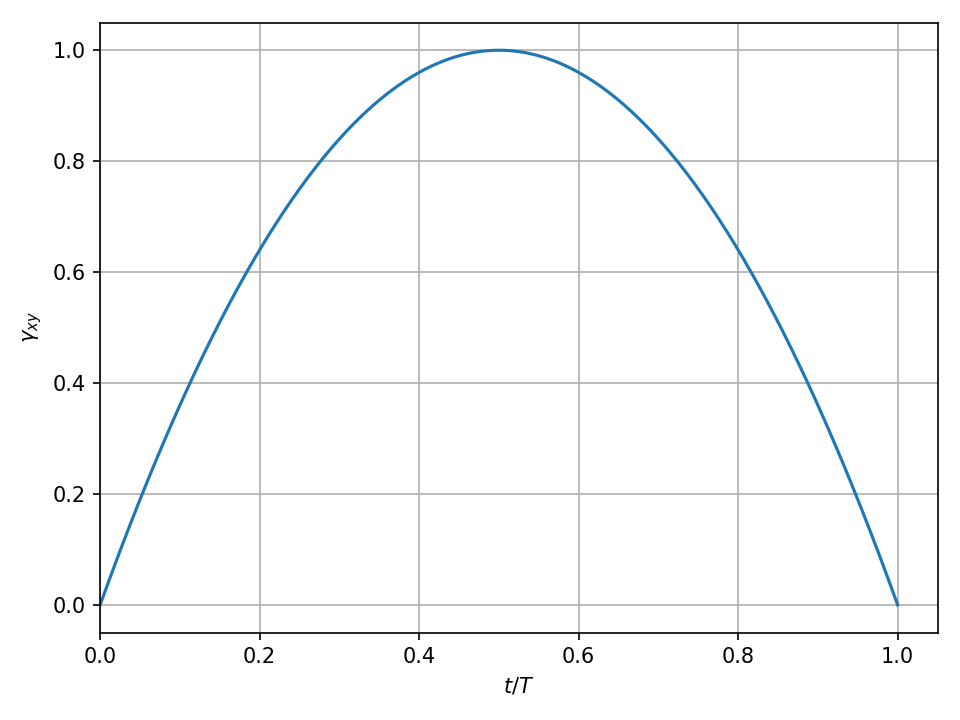}
  \includegraphics[width=0.49\textwidth]{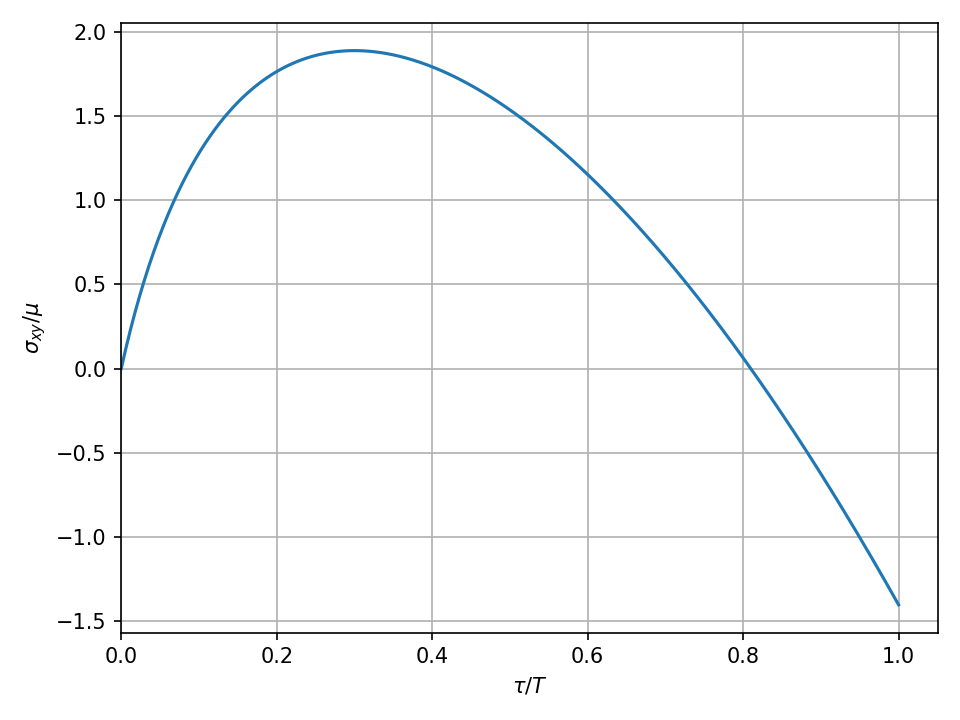}
  \caption{RVE example. Parabolic shear strain imposed on the RVE and the average stress computed on the RVE} \label{fig-poly-test2}
\end{figure}

\begin{figure}[ht]
  \centering
  \includegraphics[width=0.49\textwidth]{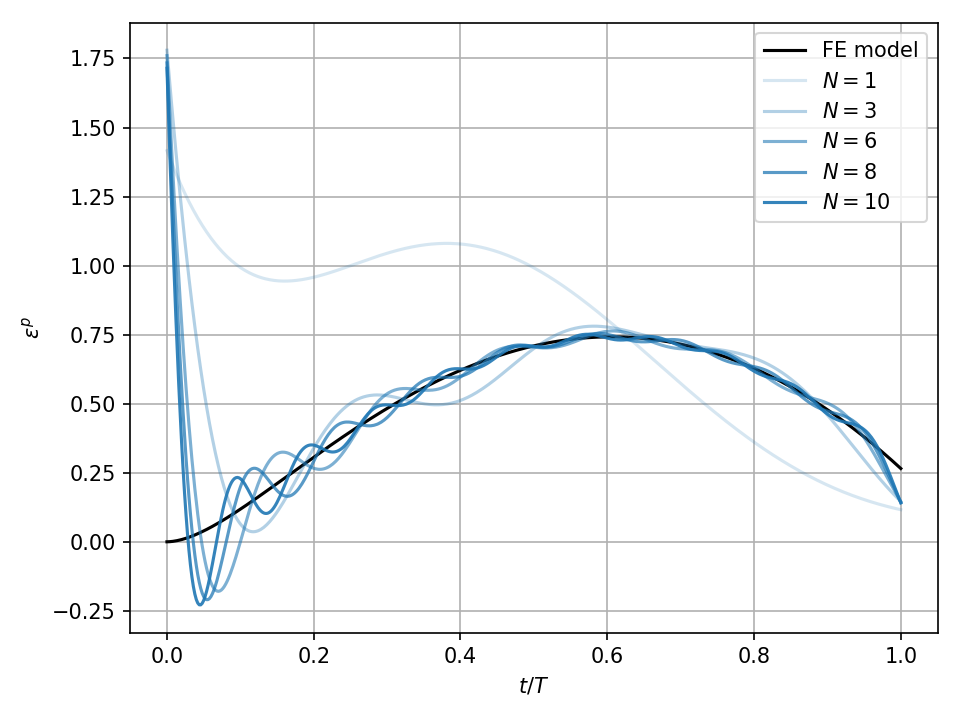}
  \includegraphics[width=0.49\textwidth]{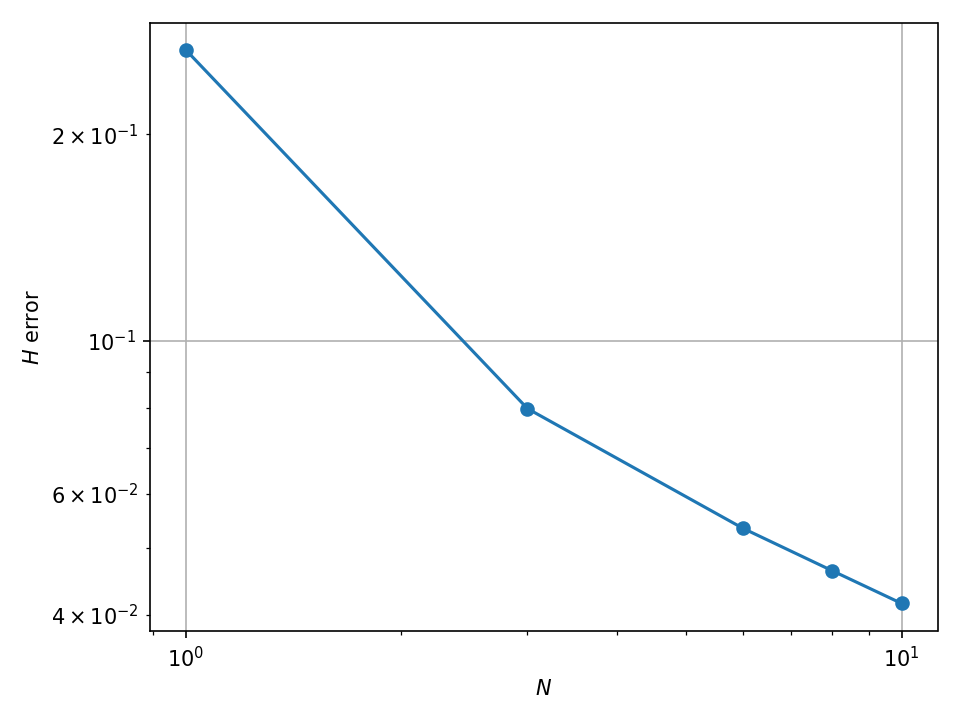}
  \caption{RVE example. Viscoelastic response of the RVE subject to shear
    strain~\eqref{eq-ex2-parabolic}. Left: inelastic strain obtained with the finite element
    model and with the optimal representation using $N$ eigenfunctions and $M=2N+1$. Right:
    $H-$norm of the error in the recovered inelastic strain.} \label{fig-ex2-convergence}
\end{figure}

We verify the convergence of the optimal rank-$N$ approximations $S_{M,N}$ by means of selected macroscopic strain histories. Specifically, we choose
\begin{equation} \label{eq-ex2-parabolic}
    \epsilon_{xy}(t)=4/25\cdot t(1-t)  
\end{equation}
and obtain the corresponding (exact) average stress evolution $\tau(t)$ directly from the finite
element model of the RVE, Fig.~\ref{fig-poly-test2}. We then compute approximate average evolutions
from $S_{M,N}$ for increasing values of $N$ and $M=2N+1$. Fig.~\ref{fig-ex2-convergence} (left) compares the evolution of the inelastic strain computed by both means, and the corresponding norm error is shown in Fig.~\ref{fig-ex2-convergence} (right). A general trend towards convergence to a floor error incurred by truncation is evident from these figures. 

\begin{figure}[ht]
  \centering
  \includegraphics[width=0.49\textwidth]{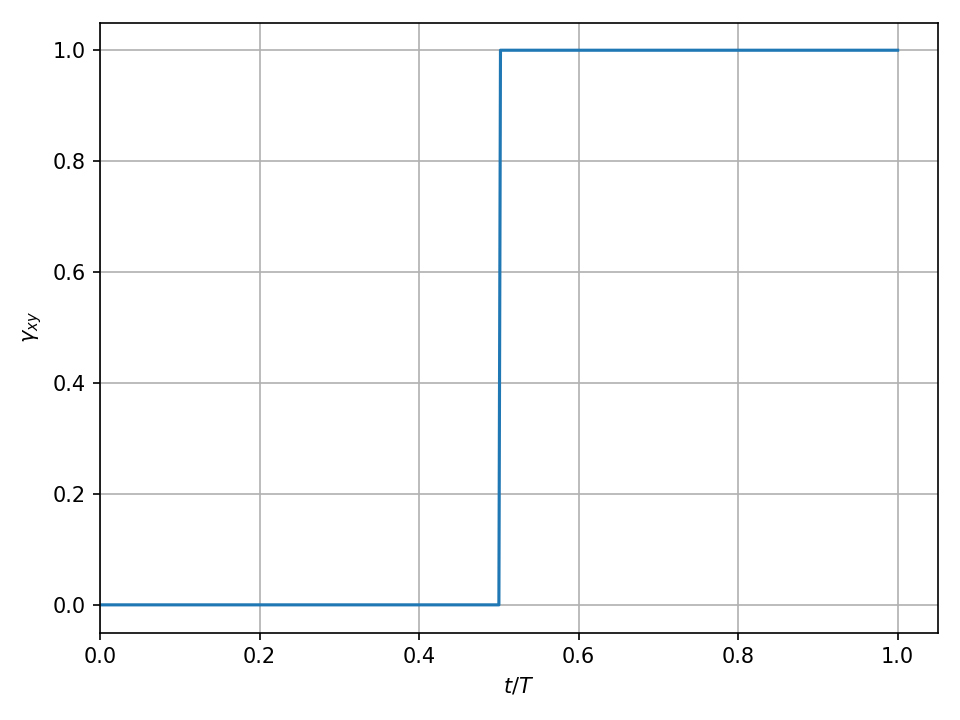}%
  \includegraphics[width=0.49\textwidth]{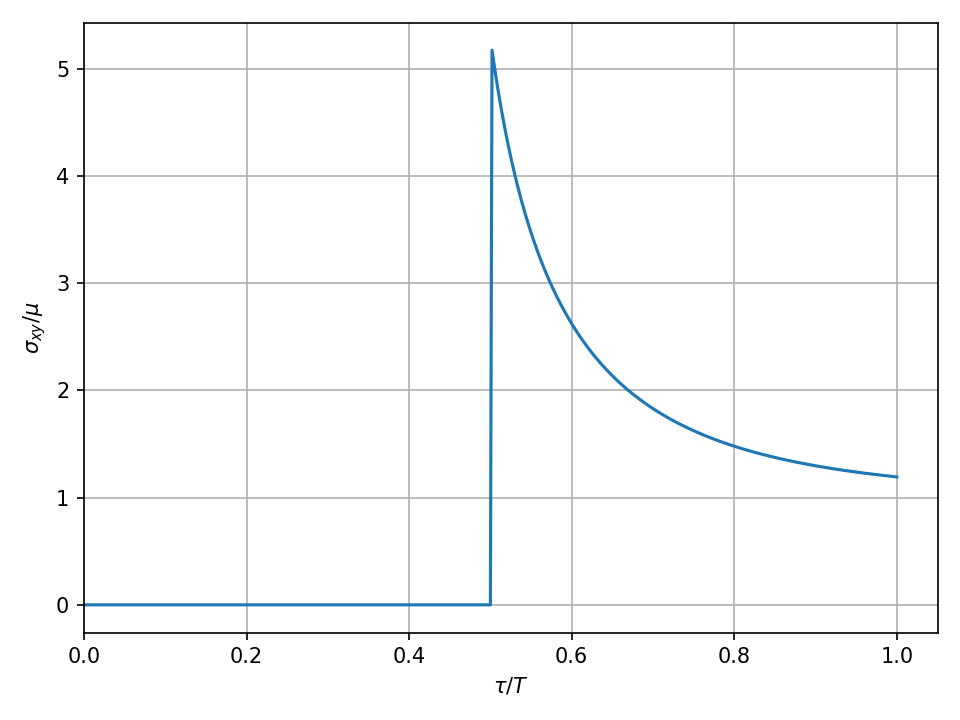}
  \caption{RVE example. Shear strain proportional to the step~\eqref{eq-ex1-step} and the  average stress computed on the RVE.} \label{fig-poly-test3}
\end{figure}

\begin{figure}[ht]
  \centering
  \includegraphics[width=0.49\textwidth]{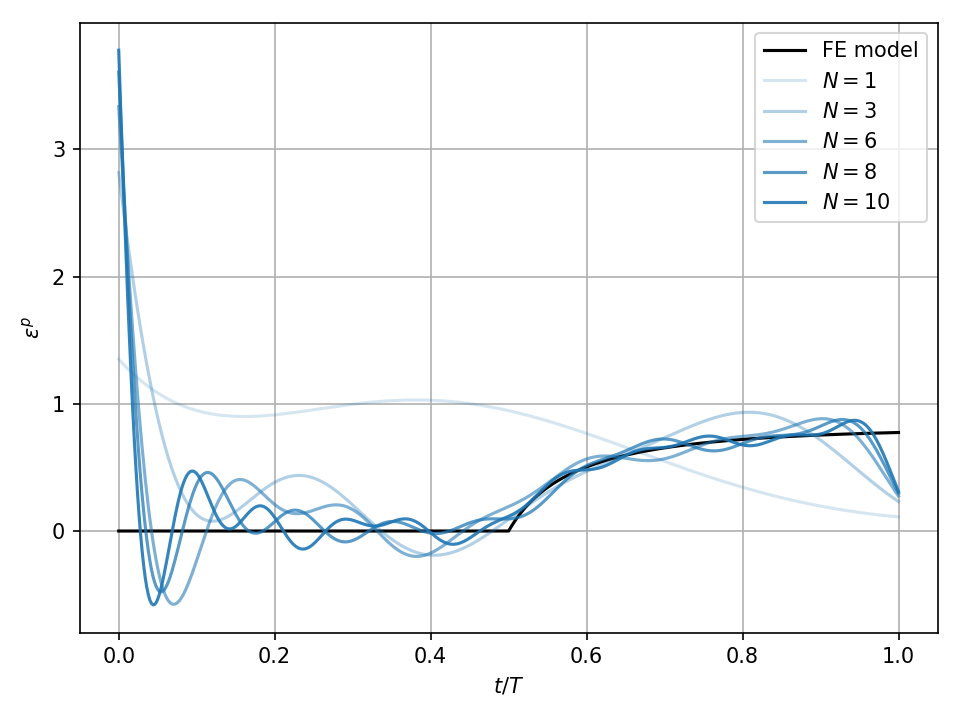}%
  \includegraphics[width=0.49\textwidth]{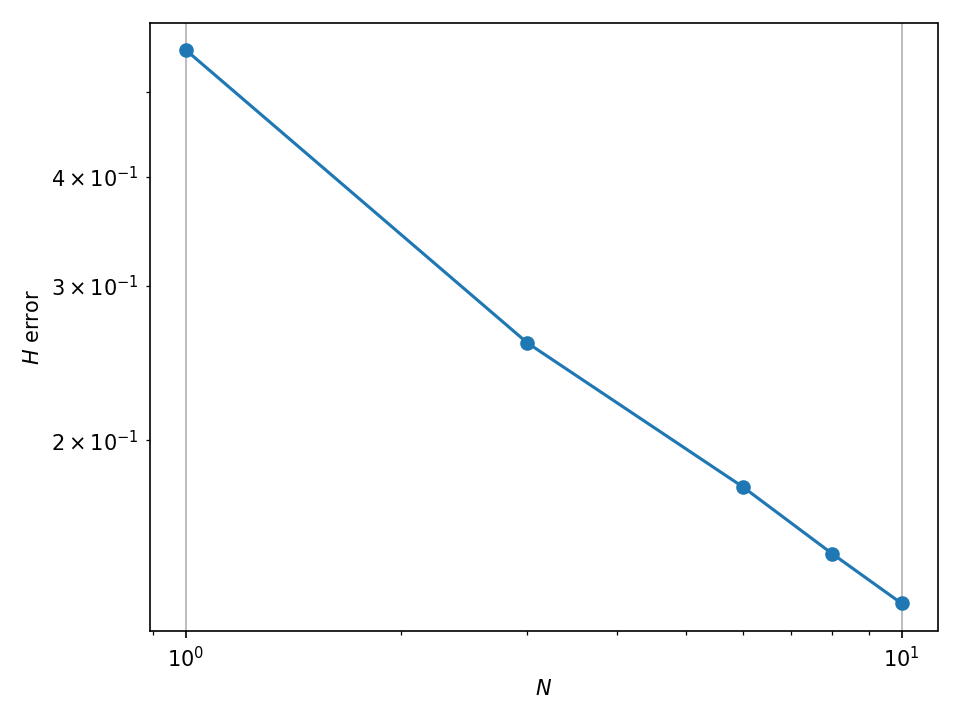}%
  \caption{RVE example. Viscoelastic response of the RVE subject 
    to shear strain~\eqref{eq-ex1-step}.
    Left: inelastic strain obtained with the finite element
    model and with the optimal representation using $N$ eigenfunctions and $M=2N+1$. Right:
    $H-$norm of the error in the recovered inelastic strain.}
  \label{fig-ex3-convergence}
\end{figure}

We repeat the same test with prescribed shear strains proportional to the step function~\eqref{eq-ex1-step}. Fig.~\ref{fig-poly-test3} shows the average shear strain and shear stress evolutions when computed by the finite element discretization. Fig.~\ref{fig-ex3-convergence} compares these exact evolutions with approximations of increasing rank and norm errors thereof. As in the preceding test case, a general trend towards convergence to a floor error incurred by truncation is evident from these figures. 

\section{Summary and Conclusions} \label{sec:conclusions}

We have developed a functional and approximation-theoretic framework for linear viscoelasticity in hereditary form, with the aim of deriving optimal low-rank representations of the hereditary functional from data. The hereditary law is expressed as a bounded Volterra operator acting on strain histories, and the elasticity tensor is used to identify dual stress--strain spaces and to define a natural Hilbert structure on the corresponding history spaces. The analysis establishes a compact operator framework for hereditary laws and characterizes optimal reduced internal-variable representations via Kolmogorov $N$-width theory. The finite-rank reduced models retain stability, thermodynamic admissibility, and possess approximation optimality. The theory clarifies the relation between hereditary constitutive structure, internal-variable representations, and low-rank approximations of the associated history operator.

Within this setting, we introduce finite-dimensional history representations and establish notions of approximation ensuring that admissible hereditary operators can be approximated arbitrarily well by finite-rank operators. We then characterize the optimal rank-$N$ approximation of the hereditary operator in operator norm and relate the resulting error to classical $N$-width quantities. These results provide, in particular, a principled criterion for selecting history variables and a quantitative estimate of the memory dimension required to attain a prescribed accuracy.

From a computational standpoint, the reduced finite-rank hereditary laws set forth efficient history representations based on a finite number of internal variables. A truncation and convergence analysis clarifies how approximation errors depend on the regularity and decay properties of the relaxation spectrum. Selected numerical tests bear out the theory and demonstrate how the proposed optimal history variables yield compact, accurate surrogates for complex rheological responses, including those arising from RVE calculations.

It bears emphasis that the optimal finite-rank representation of the hereditary law of specific materials is indifferent to the origin of the data, which can be experimental, computed from first principles, or otherwise acquired. In the present work, we have found it convenient to resort to synthetic data for purposes of demonstration, but exactly the same paradigm applies to experimental characterization of materials as well. In either case, the method calls for the stress-history response of the material to be determined for strain histories coincident with the elements of an orthonormal basis in the space of strain histories. Evidently, the choice of orthonormal basis is not unique and depends on the fading-memory properties of the material, which need to be known or surmised. However, once a proper basis is available, the question of how best to sample the material response is essentially solved, be it computationally or experimentally. 

The theory also answers the question of which are the best history variables for representing the response of a specific material. Here, by history variables we understand variables that store (partial) information about the strain history. Appealing to linearity and continuity, it is clear that history variables are simply coordinates, or linear combinations thereof, in bases spanning the space of histories. The theory of $N$-widths then identifies the optimal basis and, by extension, the optimal choice of history variables. For general hereditary laws, the approximation by finite-rank operators becomes increasingly more precise as the number of history variables is increased, and it becomes exact when the number of history variables increases to infinity. 

The present work suggests several directions for further study. Of particular interest are extensions to thermo-viscoelasticity, the treatment of nonlinear viscoelasticity and viscoplasticity via nonlinear extensions of $N$-width theory, and the integration of the present reduction strategy with data-driven identification of relaxation spectra and uncertainty quantification.

NB: A detailed {\tt Mathematica} implementation of the identification scheme and resulting viscoelastic models is provided in the supplementary materials.  

\section*{Acknowledgements}

MO gratefully acknowledges the financial support of the {\sl Centre Internacional de Mètodes Numèrics a l'Enginyeria} (CIMNE) of the {\sl Universitat Politecnica de Catalunya} (UPC), Spain, through the {\sl UNESCO Chair in Numerical Methods in Engineering}. IR acknowledges the support received from the {\sl Ministerio de Ciencia e Innovación} (Spain) under grant PLEC2023-010190.
We are also grateful for the support provided by the MORE network (Grant RED2024-153869-T) from the {\sl Ministerio de Ciencia e Innovación} (Spain).

\begin{appendix}

\section{$N$-widths}\label{app:nwidths}

For completeness we recall the connection between optimal finite-rank approximation of compact operators and the classical notion of $N$-widths; detailed treatments may be found in \cite{Tikhomirov:1960,Pinkus:1985}. Let $H_1$ and $H_2$ be Hilbert spaces and let $T\in K(H_1,H_2)$ be a compact linear operator. Denote by $T^\ast$ its adjoint. The non-negative self-adjoint operators $T^\ast T\in K(H_1,H_1)$ and $TT^\ast\in K(H_2,H_2)$ have discrete spectra that accumulate only at zero. The singular values (or $s$-numbers) of $T$ are defined by
\begin{equation}
  s_n(T)=\sqrt{\lambda_n(T^\ast T)} ,
  \quad n=1,2,\dots,
\end{equation}
where $\lambda_n(\cdot)$ denotes the eigenvalues of $T^\ast T$ arranged in non-increasing order and repeated according to multiplicity.

A fundamental result states that the best rank-$N$ approximation error of $T$ in operator norm equals $s_{N+1}(T)$. More precisely, if $\{\phi_n\}$ are orthonormal eigenvectors of $T^\ast T$ associated with $\{\lambda_n\}$ and we set $\psi_n = T\phi_n$, then the truncated singular-value expansion
\begin{equation}
  T_N x = \sum_{n=1}^N (x,\phi_n)_{H_1}\,\psi_n
\end{equation}
defines a rank-$N$ operator $T_N$ that is optimal in the sense that
\begin{equation}
  \inf_{\operatorname{rank}(A)\le N}\,\|T-A\|_{H_1\to H_2} = \|T-T_N\|_{H_1\to H_2}= s_{N+1}(T).
\end{equation}
In the body of the paper we apply this characterization with $T$ identified with the hereditary history operator, thereby obtaining optimal history variables and sharp error bounds for reduced hereditary laws.

\end{appendix}

\end{document}